\documentclass{emulateapj}
\usepackage[]{natbib, graphicx, float, appendix}
\usepackage[FIGTOPCAP]{subfigure}
\usepackage[section]{placeins}
\usepackage{wrapfig}
\usepackage[fleqn]{amsmath}
\def\vvel{\hbox{${\bf v}$}}

\newcommand{\na}{NewA}

\newcommand{\RNum}[1]{\uppercase\expandafter{\romannumeral #1\relax}}

\begin{document}
\allowdisplaybreaks
\begin{abstract}

We study a truncated accretion disk using a well-resolved, semi-global magnetohydrodynamic simulation that is evolved for many dynamical times (6096 inner disk orbits).  The spectral properties of hard state black hole binary systems and low-luminosity active galactic nuclei are regularly attributed to truncated accretion disks, but a detailed understanding of the flow dynamics is lacking.  In these systems the truncation is expected to arise through thermal instability driven by sharp changes in the radiative efficiency.  We emulate this behavior using a simple bistable cooling function with efficient and inefficient branches.  The accretion flow takes on an arrangement where a ``transition zone" exists in between hot gas in the inner most regions and a cold, Shakura $\&$ Sunyaev thin disk at larger radii.  The thin disk is embedded in an atmosphere of hot gas that is fed by a gentle outflow originating from the transition zone.  Despite the presence of hot gas in the inner disk, accretion is efficient.  Our analysis focuses on the details of the angular momentum transport, energetics, and magnetic field properties.  We find that the magnetic dynamo is suppressed in the hot, truncated inner region of the disk which lowers the effective $\alpha$-parameter by $65\%$.

\end{abstract}


\title{The Dynamics of Truncated Black Hole Accretion Disks \RNum{2}:  Magnetohydrodynamic Case}

\author{J.~Drew~Hogg\altaffilmark{1,2,3} and Christopher~S.~Reynolds\altaffilmark{1,3,4}}

\altaffiltext{1}{Department of Astronomy, University of Maryland, College Park, MD 20742, USA}
\altaffiltext{2}{NASA Earth and Space Science Fellow }
\altaffiltext{3}{Joint Space Science Institute (JSI), University of Maryland, College Park, MD 20742, USA}
\altaffiltext{4}{Institute of Astronomy, University of Cambridge, Madingley Road, Cambridge, CB3 OHA, U.K.}

\maketitle

\section{Introduction}
\label{sec-intro}

Accreting stellar mass black hole systems in black hole binaries (BHBs) display a characteristic hysteresis during outburst as they undergo state changes \citep{2004PThPS.155...99Z, 2006ARA&A..44...49R, 2007A&ARv..15....1D}.  In the low/hard state the system is less luminous and the spectrum is harder, while in the high/soft state the luminosity is dramatically increased and the spectrum is dominated by soft emission from thermal, black body emission with a peak around $\approx1$ keV.  Accreting supermassive black holes in active galactic nuclei (AGNs) seem to similarly fall into spectral classes like the BHB states, although interpreting them as true analogs is treacherous because the long evolutionary time scales prevents the direct observation of any possible hysteresis cycle.  Low-luminosity AGNs (LLAGNs) are observed to have low accretion rates, low luminosities, and hard spectra which is interpreted as evidence of accretion through a hot, diffuse flow \citep{1982Natur.295...17R, 1995ApJ...444..231N, 2008ARA&A..46..475H, 2014ARA&A..52..529Y}.  The standard Seyfert or quasar, on the other hand, has a higher accretion rate, higher luminosity, and a softer spectra dominated by a thermal continuum which peaks in the ultraviolet.

The canonical model behind the two accretion state paradigm in BHBs and AGNs attributes the distinct properties to varying degrees of disk truncation \citep{1997ApJ...489..865E, 2011ApJ...733...60T, 2011ApJ...726...87Y, 2014MNRAS.438.2804N}.  The putative truncation is believed to arise as the gas transitions from a radiatively-efficient, geometrically-thin flow in the outer regions of the disk to a radiately-inefficient, geometrically-thick flow in the inner region.  When the disk is truncated and the inner region of the disk is occupied by gas in the hot phase, the system should be less luminous and have a harder spectrum as free-free emission and Comptonization dominates the radiation.  When the optically-thick, geometrically-thin accretion disk extends down to the innermost stable circular orbit (ISCO), the system should be more luminous and the spectrum should be dominated by quasi-blackbody thermal radiation.

Additional evidence for disk truncation in BHBs in the high/soft state and LLAGNs comes from X-ray monitoring of the 6.4 keV Fe $K\alpha$ line.  In BHB systems like GX 339-4 \citep{2009ApJ...707L..87T} and H1743-322 \citep{2014ApJ...789..100S}, the line profile is seen to evolve as the outburst evolves.  In the low/hard state, the line is narrower and best fit if the line originates from large radii, while in the high/soft state, the line is broader and best fit if the bulk of the emission originates close to the central black hole.  Narrow Fe K$\alpha$ lines are also observed from LLAGNs and associated with a truncated accretion disk in systems like M81 \citep{2004ApJ...607..788D}, NGC 4579 \citep{2004ApJ...607..788D}, NGC 4593 \citep{2004MNRAS.352..205R, 2007ApJ...666..817B, 2009ApJ...705..496M}, and NGC 4258 \citep{2009ApJ...691.1159R}.  This seemingly direct evidence of truncation lends credibility to the model, but it is not quite so clear.  In fact, broadened Fe $K\alpha$ lines from relativistic blurring have been reported in a number of sources when they are in the (luminous) hard state, including GRS 1739-278 \citep{2015ApJ...799L...6M}, Cygnus X-1 \citep{2015ApJ...808....9P} and even GX 339-4 \citep{2006ApJ...653..525M}, suggesting the optically-thick disk extends to the ISCO even in the hard state.

A transition in the radiative efficiency makes it particularly challenging to understand the flow dynamics.  Separately, much effort has gone into studying cooler accretion flows in radiatively-efficient, geometrically-thin disks and hot flows in non-radiative, geometrically-thick disks.  However, merging the two models is not straightforward and has a strong dependence on whether the disk is considered in the hydrodynamic (HD) or magnetohydrodynamic (MHD) regime.  

The seminal work of \citet{1973A&A....24..337S} provided the framework for developing thin disk theory by parameterizing the anomalous viscosity with a dimensionless ``$\alpha$"-parameter.  The magnetorotational instability \citep[MRI;][]{Velikhov59, 1960PNAS...46..253C, 1991ApJ...376..214B} has since been accepted as the most likely driver of the vigorous disk turbulence that provides the angular momentum transport.  Numerically, thin disks have been studied with varying degrees of sophistication in unstratified shearing boxes \citep[e.g.][]{1991ApJ...376..223H, 1995ApJ...440..742H}, stratified shearing boxes \citep[e.g.][]{1995ApJ...446..741B,1996ApJ...463..656S}, unstratified global models \citep[e.g.][]{1998ApJ...501L.189A, 2001ApJ...554..534H, 2001ApJ...548..868A, 2003MNRAS.341.1041A, 2012ApJ...749..189S}, stratified global disk models \citep[e.g.][]{2000ApJ...528..462H, 2001ApJ...548..348H, 2009ApJ...692..869R, 2010ApJ...712.1241S}, relativistic global models \citep[e.g.][]{2003ApJ...599.1238D, 2003ApJ...589..444G, 2006MNRAS.368.1561M}, and global models with radiative transfer \citep[e.g.][]{2014ApJ...796..106J,2014MNRAS.441.3177M, 2015MNRAS.447...49S}.  In general terms, agreement has been reached that the gas motions are dominated by anisotropic MHD turbulence and that angular momentum is transported through stresses arising from correlated fluctuations in the fluid velocities, the Reynolds stress $R_{r \phi} = \rho v_{R} \delta v_{\phi}$, and correlated fluctuations in the magnetic field, the Maxwell stress $M_{r \phi} = -B_{r} B_{\phi}/4\pi$ \citep{1994MNRAS.271..197B}.  The $\alpha$-parameter is thus the ratio of the shell averaged total stress to the shell averaged pressure, \begin{equation}
\label{eqn-alpha}
\alpha=\frac{\langle M_{r\phi} + R_{r\phi}\rangle}{\langle P \rangle}.
\end{equation}

A complete understanding of hot accretion flows has remained elusive.  Often, radiatively inefficient accretion flows are considered to be advection dominated accretion flows \citep[ADAFs;][]{1994ApJ...428L..13N, 1995ApJ...444..231N, 1995ApJ...452..710N}.  However, entropy gradients in the ADAF imply that the flow is unstable to convection according to the Solberg-H{\o}iland criterion, leading to a convection dominated accretion flow \citep[CDAF;][]{2000ApJ...539..809Q}.  Indeed, HD \citep{1996ApJ...464..364S, 1999MNRAS.303..309I, 2000ApJ...537L..27I, 2003ApJ...582...69P, 2010MNRAS.408.1051Y, 2012ApJ...761..130Y} and MHD \citep{2001MNRAS.322..461S, 2001ApJ...554L..49H, 2002ApJ...573..738H, 2003ApJ...592.1042I, 2012MNRAS.426.3241N} models of hot accretion flows have displayed circulation within the flow, resembling convective eddies.  The most current MHD models find this apparent convective motion is stabilized, though \citep[e.g.][]{2012ApJ...761..130Y, 2012MNRAS.426.3241N, 2015ApJ...804..101Y}.

Computational and algorithmic advances have recently enabled the numerical exploration of BHB state transitions.  HD models \citep{2013MNRAS.435.2431D, 2016MNRAS.tmp..530W} which self-consistently account for radiation from bremsstrahlung, synchrotron and synchrotron self-Comptonization demonstrate one of the basic tenets of the current state transition model: that a cold, dense disk forms out of the hot, diffuse medium once the accretion rate exceeds a critical threshold.  As the accretion rate and density increase, cooling becomes more efficient in the inner region of the disk, thereby decreasing the radius of the truncation.  Using a three-dimensional MHD simulation which accounts for general relativity and radiation, \citet{2016ApJ...826...23T} corroborate this finding.  

Despite the success simulations have had in demonstrating the viability of the truncated disk model to explain the phenomenology observed from BHB and AGN systems, they have yet to yield a detailed understanding of the gas dynamics and disk evolution in a truncated disk system.  Properly modeling a truncated disk requires a simulation grid capable of resolving the internal dynamics of the thin-disk, a large computational volume, and a long enough integration that global processes can develop over several viscous timescales.  Fundamentally, these requirements are at odds with each other, requiring sacrifices in various aspects of the simulation.  This paper is a continuation of the work of \citet{2017ApJ...843...80H} (Paper I) where we implemented an \emph{ad hoc} bistable cooling prescription to force a truncation in a two-dimensional, axisymmetric, HD, viscous accretion disk.  Here, we extended the model to a three-dimensional, MHD accretion disk, which better represents a physical system since MRI driven turbulence provides an effective viscosity and heats the gas.  

In both of these simulations, we prioritize resolution and evolutionary time over less influential physics.  For instance, we simulate both disks with non-relativistic physics in a Newtonian potential since empirical measurements place the truncation at large radii where GR effects should be negligible.  Additionally, the cooling function used in these simulations is computationally simple, but still captures the essence of the true radiative physics, albeit in an approximate manner.  

The HD case in Paper I provides a benchmark against which we can compare the behavior of the MHD accretion disk.  In Paper \RNum{1}, several HD features significantly affect the evolution.  The launching of buoyant bubbles within the flow as gas transitioned from the efficient cooling branch to the inefficient cooling branch was crucial in shaping the global disk dynamics.  This cyclical process involved the interplay of the gas density and temperature, and introduced time variability to the angular momentum transport efficiency and energetics.  Buoyant bubbles drove an outflow, leading to a circulation pattern in the disk atmosphere.  Gas was expelled from the inner disk, but fell back and was eventually accreted.  Additionally, a thin stream of cooler gas reached from the disk body into the truncated region, almost to the inner boundary.  Since we considered a full viscous prescription with an isotropic viscous stress tensor, the $\theta-\phi$ stress component was influential where vertical velocity shears were present, for instance between the Keplerian disk body and the sub-Keplerian outflow in the disk atmosphere.

Considering a MHD accretion disk introduces additional complexity, most notably the role of turbulence and the large-scale magnetic dynamo.  The chaotic disk turbulence, by its nature, acts to disrupt coherent structure on small scales.  Much of the behavior in the HD accretion disk that developed on large scales originated on much smaller scales.  The large-scale magnetic dynamo naturally induces low-frequency variability in the accretion disk.  Dynamos, in some form, seem to be a natural occurrence in rotating plasmas.  While magnetic activity cycles with long periods have been known to exist in stars for many decades \citep{1978ApJ...226..379W}, we have only just begun to uncover the role dynamos play in moderating the evolution and time variability of black hole accretion disks.  Numerically, dynamos are ubiquitously observed in shearing box and global simulations of MRI driven turbulence.  The large-scale dynamo is characterized by oscillation of the global magnetic field orientation and amplitude.  Except for extreme circumstances, like a strong magnetic field \citep{2016MNRAS.457..857S}, the dynamo frequency is regularly found to be $\Omega_d\approx5-20 \: \Omega_K$, where $\Omega_K$ is the orbital frequency.  This becomes important because the dynamo modulates the local Maxwell stress \citep{2010ApJ...713...52D, 2012ApJ...744..144F, 2016ApJ...826...40H}, which consequently influences variability in the angular momentum transport and, presumably, gas heating.

In this paper we begin with a summary of the numerical model in Section \ref{sec-model} and present an overview of the initial setup, bistable cooling function, a test simulation and a resolution assessment.  In Section \ref{sec-results} we present our analysis of the truncated disk's dynamics, angular momentum transport, energetics, and magnetic field properties.  We summarize our analysis and discuss the implications of our results in Section \ref{sec-discussion} and provide closing remarks in Section \ref{sec-conclusion}.

\section{Numerical Model}
\label{sec-model}

In this paper we consider a three-dimensional, semi-global accretion disk with a bistable cooling curve.  The disk was constructed to match the HD model of Paper \RNum{1}, but extended to MHD and three dimensions.  An overview of the bistable cooling function we used is given here, as a more rigorous discussion and motivation is presented in Paper I.  The cooling prescription is calibrated through a pre-chosen effective $\alpha$ parameter, which we determined with a calibration simulation.  The fiducial model is evolved in two stages.  The first stage is an initialization where the model is evolved with a constant disk aspect ratio ($h/r=0.1$) for $1.99\times10^5\:GM/c^3$ to allow the MRI driven turbulence to fully develop in the simulation volume.  At the inner-edge of the simulated disk ($r=10\:r_g$), this is 1000 orbital timescales where $t_{orb,10}=199\:GM/c^3$.  Then the cooling function is switched to the bistable cooling function described below and the model evolves for an additional $1.99\times10^5\:GM/c^3$ to allow the truncated disk to fully develop and the unphysical transient behavior to die away before we begin our analysis.  The second stage is the portion of the simulation used in our analysis and lasts $t=8.14\times10^5\:GM/c^3$ (4096 $t_{orb,10}$).

\subsection{Simulation Code}

We use the second-order accurate PLUTO \emph{v4.2} code \citep{2007ApJS..170..228M} to solve the equations of ideal MHD, \begin{eqnarray}
\frac{\partial \rho}{\partial t}+\nabla\cdot(\rho\vvel)&=&0,\\
\frac{\partial}{\partial t}(\rho\vvel)+\nabla\cdot(\rho\vvel\vvel-{\bf BB}+P{\cal I})&=&-\rho\nabla\Phi, \qquad \\
\frac{\partial}{\partial t}(E+\rho\Phi)+\nabla\cdot\left[(E+P+\rho\Phi)\vvel-\bf{B}(\vvel\cdot\bf{B})\right]&=&-\Lambda,\\
\frac{\partial{\bf{B}}}{\partial{t}}={\bf\nabla\times}({\bf{v}\times{B}}),
\end{eqnarray}
where $\rho$ is the gas density, $\vvel$ is the fluid velocity, $P$ is the gas pressure, {\boldmath$B$} is the magnetic field, ${\cal I}$ is the unit rank-two tensor, $E$ is the total energy density of the fluid,
\begin{equation}
E=u+{1\over 2}\rho |\vvel|^2 +\frac{\bf{B}^2}{2},
\end{equation}  
and $\Lambda$ accounts for radiative losses through cooling.  The equations were solved using the {\tt hlld} Riemann solver with the code in the dimensionally unspilt mode.  A piecewise parabolic reconstruction is used in space and time integration is done with a second-order Runge Kutta algorithm.  PLUTO is a Godunov code, so the total energy in the simulation is conserved except for losses across the boundary and energy removed through our cooling function.  The method of constrained transport is used to enforce the $\nabla \cdot {\bf B} = 0$ condition.

\subsection{Simulation Setup}
\label{sec-sim_setup}

The simulation grid is the same as in Paper I, but expanded in the $\phi$-direction, rather than being treated axisymmetrically.  Spherical coordinates are used in a computational domain that spans $r \in [10 r_{g}, 1000 r_{g}]$, $\theta \in [\pi/2-1.0, \pi/2+1.0]$, and  $\phi \in [0.0, \pi/2]$ with $N_R \times N_\theta \times N_\phi  = 512 \times 512 \times 128= 3.35 \times 10^7$ zones.  Logarithmic spacing is used in the radial coordinate so that $\Delta r / r$ is constant.  In the $\theta$-direction, the simulation uses 320 uniformly grid zones in the midplane region from $\theta=\pi/2\pm0.5$ and 96 geometrically stretched grid zones in each of the coronal regions (i.e. $\pi/2-1.0 \leq \theta \leq \pi/2-0.5$ and $\pi/2+0.5 \leq \theta \leq \pi/2+1.0$).  Outflow is allowed through the $r$ and $\theta$ boundaries while the the $\phi$ boundaries are periodic.  To prevent artificially low or negative values, floors are imposed on density and pressure.

Since the truncation is expected to occur at $30-100\:r_g$ \citep{2014MNRAS.438.2804N}, we can consider a scale free Newtonian potential, \begin{equation}
\Phi=-\frac{GM}{r}.
\end{equation} 
Similarly, all fluid variables (e.g. $\rho$, $p$, $T$) are considered and reported in a normalized, scale free form.  The gas is considered ideal with a $\gamma = 5/3$ adiabatic equation of state.

To initialize the simulation, we use the steady state $\alpha$-disk solution: \begin{equation}
\rho(R, \theta)=\rho_{0} R^{-3/2}\exp\Bigg(-\frac{z^2}{2 c_{s}^2 R^3}\Bigg)\Bigg[ 1-\Bigg(\frac{R_{*}}{R}\Bigg)^{\frac{1}{2}}\Bigg]
\end{equation} where $\rho_{0}$ is a normalization constant to set the maximum disk density to unity, $R$ is the cylindrical radius ($R=r \cos\theta$), $z$ is the vertical disk height ($z=R\sin\theta$), and $c_{s}$ is the isothermal sound speed ($c_{s}=\sqrt{P/\rho}=h v_{K}$).  $R_{*}$ accounts for the torque-free inner boundary condition and is set just inside the inner simulation domain ($R_{*}=11\:GM/c^2$).

The gas starts with a purely azimuthal velocity that is set such that the effective centripetal force balances the gravitational force.  The initial magnetic field is weak ($\langle 8\pi p_{gas}/B^2\rangle=200$) and set from a vector potential,
\begin{eqnarray}
&A_r=0, \\
&A_{\theta}=0, \\
&A_{\phi} = A_0 p^{1\over2} e^{-(z/h)^4} R \sin\bigg( \frac{\pi \log(R/2)}{h}\bigg),
\end{eqnarray}
where $A_0$ is a normalization constant.  Beyond $|z|=2.5 h$ and within $r=15\:r_g$, $A_{\phi}$ is set to zero.  Because of these field loops, the gas is unstable to the MRI.

To help manage the data volume, simulation data was dumped every 2 $t_{orb,10}$ ($\Delta t = 397.4 \: GM/c^3$).

\subsection{Test Simulation}
\label{sec-test_sim}

The bistable cooling function used in the simulation assumes a balance between the global heating and cooling and requires an effective $\alpha$-parameter be known \emph{a priori}.  To determine this value, we ran a test simulation using the same grid and initial conditions that are used in our fiducial MHD model.  Instead of the bistable cooling function, a simple \citet{2009ApJ...692..411N} style cooling function enforces a target scale height ratio, $h/r=0.1$ in this case, such that $P_{targ}=\rho v_{\phi}^2 (h/r)^2/\gamma$.  In terms of pressure the cooling function is, $\Lambda = f(P-P_{targ})/\tau_{cool}$, where $f$ is a switch function ($f=0.5[(P-P_{targ})/|P-P_{targ}|+1]$) and $\tau_{cool}$ is the cooling time.  We take $\tau_{cool}$ to be the orbital time to strictly enforce the scale height ratio.  

The simulation was run for $t=1.01\times10^{5}\:GM/c^3$ (512 $t_{orb,10}$) to allow the MRI driven turbulence to fully develop and saturate in the inner third of the simulation domain.  Averaged over the final 50 orbits of the simulation, we find $\langle \alpha \rangle=0.065$ which we use to set the cooling rates of the bistable HD and MHD simulations.

\subsection{Cooling Function}

Our overarching goal is to characterize the dynamics of a truncated accretion disk.  To this end, a thermodynamic transition between an efficient cooling state and an inefficient cooling state must arise at some radius to produce the truncation.  We accomplish this by treating the gas cooling with an \emph{ad hoc} optically-thin cooling function with two branches reflecting the efficient and inefficient cooling, a similar technique to that employed by \citet{2012MNRAS.424..524F} to study how accretion disk geometry influences jet power.  Starting from the simple, viscous formalism of \citet{1973A&A....24..337S}, the volumetric heating rate is \begin{equation}
{\cal H}={3\over 2}\alpha{\cal R}\rho T\Omega_K,
\end{equation} where ${\cal R}$ is the gas constant.  In a steady-state disk, ${\cal H} = \Lambda$.  Taking as an \emph{ansatz} that the cooling function has the form, \begin{equation}
\Lambda=\Lambda_0(T)\rho T^2f(r),
\end{equation} we can construct a cooling law that is positive-definite everywhere, viscously stable, and produces a geometrically-thick solution at small radii and a geometrically-thin solution at large radii.

We seek to cool to a local equilibrium temperature.  This is related to a target scale height ratio by $c_s^2={\cal R}T=(h/r)^2\Omega_K^2$.  The ${\cal H}=\Lambda$ condition gives \begin{equation}
\frac{h}{r}=\sqrt{\frac{3\alpha{\cal R}^2}{2\Lambda_0(T)r^2\Omega_Kf(r)}}.
\end{equation}  For convenience, we take $f(r)={\cal R}^2r^{-2}\Omega_K^{-1} \propto r^{-1/2}$ so that \begin{equation}
\frac{h}{r}=\sqrt{\frac{3\alpha}{2\Lambda_0(T)}}.
\end{equation} The disk will have a constant aspect ratio if $\Lambda_0$ is constant.  If there is an abrupt decrease in $\Lambda_0(T)$ at a prescribed temperature $T_T$, \begin{eqnarray}
\Lambda_0(T)=\begin{cases}
\Lambda_c \hspace{1cm}(T<T_T)\\
\Lambda_h \hspace{1cm}(T\ge T_T)\\
\end{cases}
\end{eqnarray} where $\Lambda_c$, $\Lambda_h$ are constants and $\Lambda_c > \Lambda_h$, we can induce a thermal instability and consequently a radial transition into the flow.

In this formulation, the ${\cal H}=\Lambda$ condition dictates three regions exist in our disk.  An inner truncation radius occurs at $r_{tr,i}=3\alpha GM/2\Lambda_c{\cal R}^2T_T$ inside of which thermal balance requires the flow to be hot with geometric thickness\begin{equation}
\left(\frac{h}{r}\right)_{\rm hot}=\sqrt{\frac{3\alpha}{2\Lambda_h}}\hspace{0.5cm}{\rm for}\hspace{0.5cm}r<r_{tr,i}:=\frac{3\alpha GM}{\Lambda_c{\cal R}^2T_T}
\vspace{0.01cm}
\end{equation} Outside of $r_{tr,o}=3\alpha GM/2\Lambda_h{\cal R}^2T_T$ the flow will be cold with geometric thickness \begin{equation}
\left(\frac{h}{r}\right)_{\rm cold}=\sqrt{\frac{3\alpha}{2\Lambda_c}}\hspace{0.5cm}{\rm for}\hspace{0.5cm}r>r_{tr,o}:=\frac{3\alpha GM}{\Lambda_h{\cal R}^2T_T}
\end{equation}  The flow is bistable in between these two radii and thermal balance is satisfied on either the hot or cold branch with geometric scale height \begin{equation}
\left(\frac{h}{r}\right)_{\rm cold} \hspace{0.2cm}{\rm or}\hspace{0.3cm}\left(\frac{h}{r}\right)_{\rm hot} \hspace{0.3cm}{\rm for}\hspace{0.3cm}r_{tr,i}<r<r_{tr,o}.
\end{equation}

The ratio of the cold/hot geometric thickness is related to the ratio of the inner/outer transition radii,
\begin{equation}
\frac{r_{tr,o}}{r_{tr,i}}=\left[\frac{(h/r)_{\rm hot}}{(h/r)_{\rm cold}}\right]^2
\end{equation}

In this work, we set
\begin{eqnarray}
&\left(\frac{h}{r}\right)&_{\rm cold}=0.1\\
&\left(\frac{h}{r}\right)&_{\rm hot}=0.4\\
&r_{tr,i}&=100r_g.
\end{eqnarray}
which implies $r_{tr,o}=1600r_g$.  As we shall see, this toy cooling function results in a successful transition from a cold/thin to a hot/thicker disk close to the inner transition radius. Interestingly, the flow in the bistable region remains predominantly in the cold/thin state.

\subsection{Resolution Diagnostics}
\label{sec-resolution}

When interpreting results from numerical MHD accretion disk models, the issues of convergence and resolvability are paramount.  As a result, much effort has gone into developing criteria and diagnostics to judge the fidelity of these models.  Here, we apply a standard suite of tools to demonstrate that this model is adequately resolved.  All measurements are averaged over time and volume.  Time averages are taken during the first initialization stage before the change in the cooling law from $t=1.79-1.99\times10^{5}\:GM/c^3$  (900-1000 $t_{orb,10}$).  Volume averages are taken within the main disk body ($h/r<0.1$) and within the region where the MRI turbulence has saturated ($r<500\:r_g$) during this time.  These can be taken as lower bounds on the resolution because the disk inflates in the analysis stage of the simulation to $h/r>0.1$ and the effective disk resolution increases.

At the minimum target scale height ratio of $h/r=0.1$, the simulation grid covers one disk scale height with 32 zones.  This value has been shown to be a crude threshold where the macroscopic disk behavior and simulation results converge \citep{2012ApJ...749..189S, 2013ApJ...772..102H}.  Quality factors provide a way to assess resolvability of the simulation by comparing the characteristic MRI wavelength in each direction, $\lambda_{MRI}=2\pi v_A /\Omega$, where $v_A$ is the Alfv\'en speed, to the grid zone size, i.e. the $\theta$ component in the vertical direction and $\phi$ component in the azimuthal direction \citep{2011ApJ...738...84H}.  Strictly speaking, they only measure the linear MRI amplitude, but they serve as a probe of the fully nonlinear turbulence.  The quality factors are, \begin{equation}
Q_\theta=\frac{\lambda_{MRI,\theta}}{R\Delta\theta}
\end{equation}
and \begin{equation}
Q_\phi=\frac{\lambda_{MRI,\phi}}{R\Delta\phi}.
\end{equation}
We find that $\langle Q_\theta \rangle = 16.5$ and $\langle Q_\phi \rangle = 19.5$.  This is considered``well-resolved" as $Q_\theta>6-8$ has been shown to properly capture the growth rate of the MRI \citep{1995CoPhC..89..127H, 2004ApJ...605..321S, 2010A&A...516A..26F}.  The resolvability criteria for capturing the nonlinear behavior of the MRI driven turbulence is somewhat stricter, $Q_\theta>10$ and $Q_\phi>20$ \citep{2011ApJ...738...84H, 2012ApJ...749..189S, 2013ApJ...772..102H}, although they are not independent of each other.  Nevertheless, we either meet or exceed these guidelines.

\begin{figure}
\includegraphics{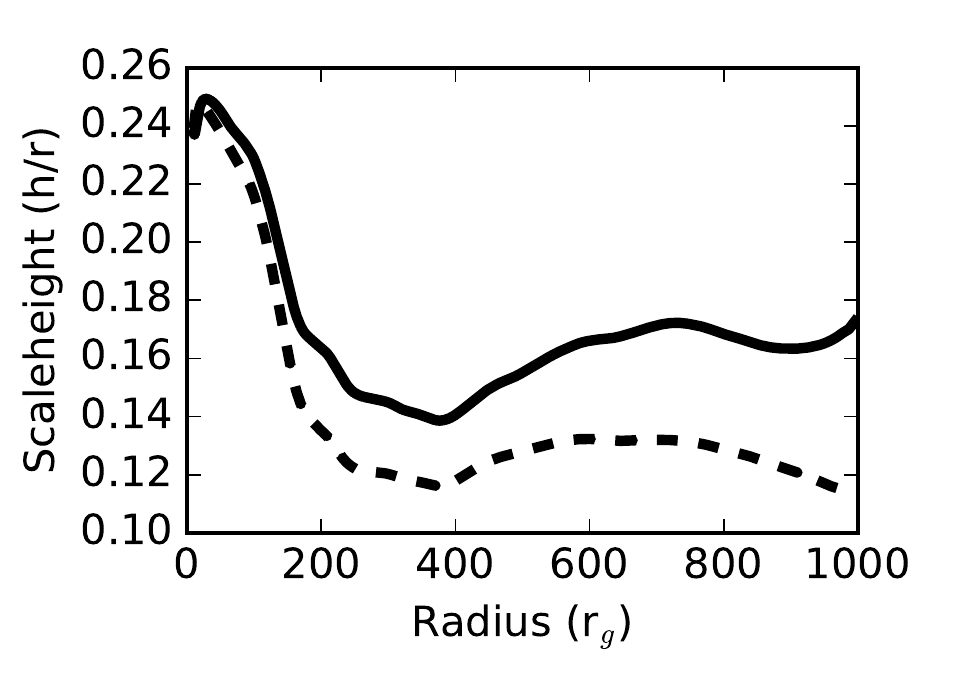}
\caption{Average $h_{g}(r)/r$ (solid line) and $h_{T}(r)/r$ (dashed line).  Averaging has been done over time.
\label{fig-scaleheights}}
\end{figure}

The saturation level of the anisotropic turbulence can be measured through the characteristic orientation of the magnetic field, given by the average in-plane magnetic tilt angle, \begin{equation}
\Theta_B = -\arctan\bigg(\bigg\langle \frac{B_r}{B_\phi} \bigg\rangle\bigg).
\end{equation}
Theoretical estimates predict $\Theta_B\approx15^{\circ}$ \citep{2009ApJ...694.1010G, 2010ApJ...716.1012P} and it is empirically found to be between $\Theta_B\approx11-13^{\circ}$ in high resolution simulations of local \citep{2011ApJ...738...84H} and global \citep{2012ApJ...749..189S, 2013ApJ...772..102H, 2016ApJ...826...40H} MRI driven turbulence.  We measure $\langle\Theta_B\rangle=12.3^{\circ}$, which is within the bounds of previously measured saturated turbulence. 

\section{Results}
\label{sec-results}

In our analysis we are primarily focused on the dynamics of the truncated disk and any sources of variability that could potentially be observable.  We strive to understand the drivers behind the disk behaviors and aim to dissect the details of the angular momentum transport and disk energetics.  To this end, we take a broad approach and consider all aspects of the accretion flow by investigating how properties like the gas temperature, motions, magnetic field strength, and density vary in both space and time and their interdependencies.  The analysis is specifically interested in understanding how the presence of the truncation alters the components of the accretion flow interior to, in, and beyond the truncation.  In many cases the analysis considers both the average and instantaneous properties to develop the most complete understand of the truncated disk behavior.

\subsection{The Truncated Disk}

\begin{figure*}[t]
  \subfigure[$\rho$ at $t=1.99\times10^{5}\:GM/c^3$]{\includegraphics[width=0.5\textwidth]{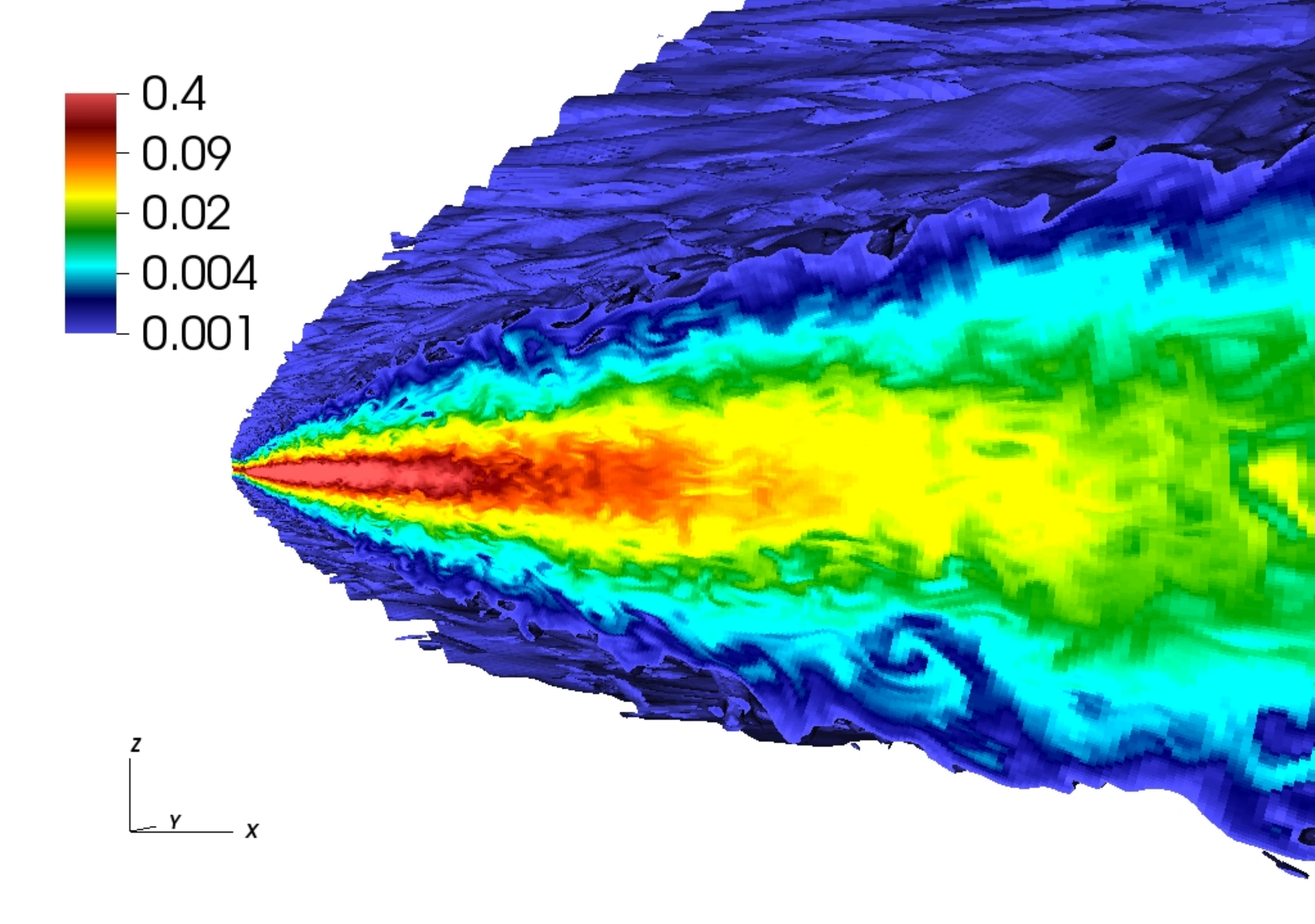}}
  \subfigure[$h_T/r$ at $t=1.99\times10^{5}\:GM/c^3$]{\includegraphics[width=0.5\textwidth]{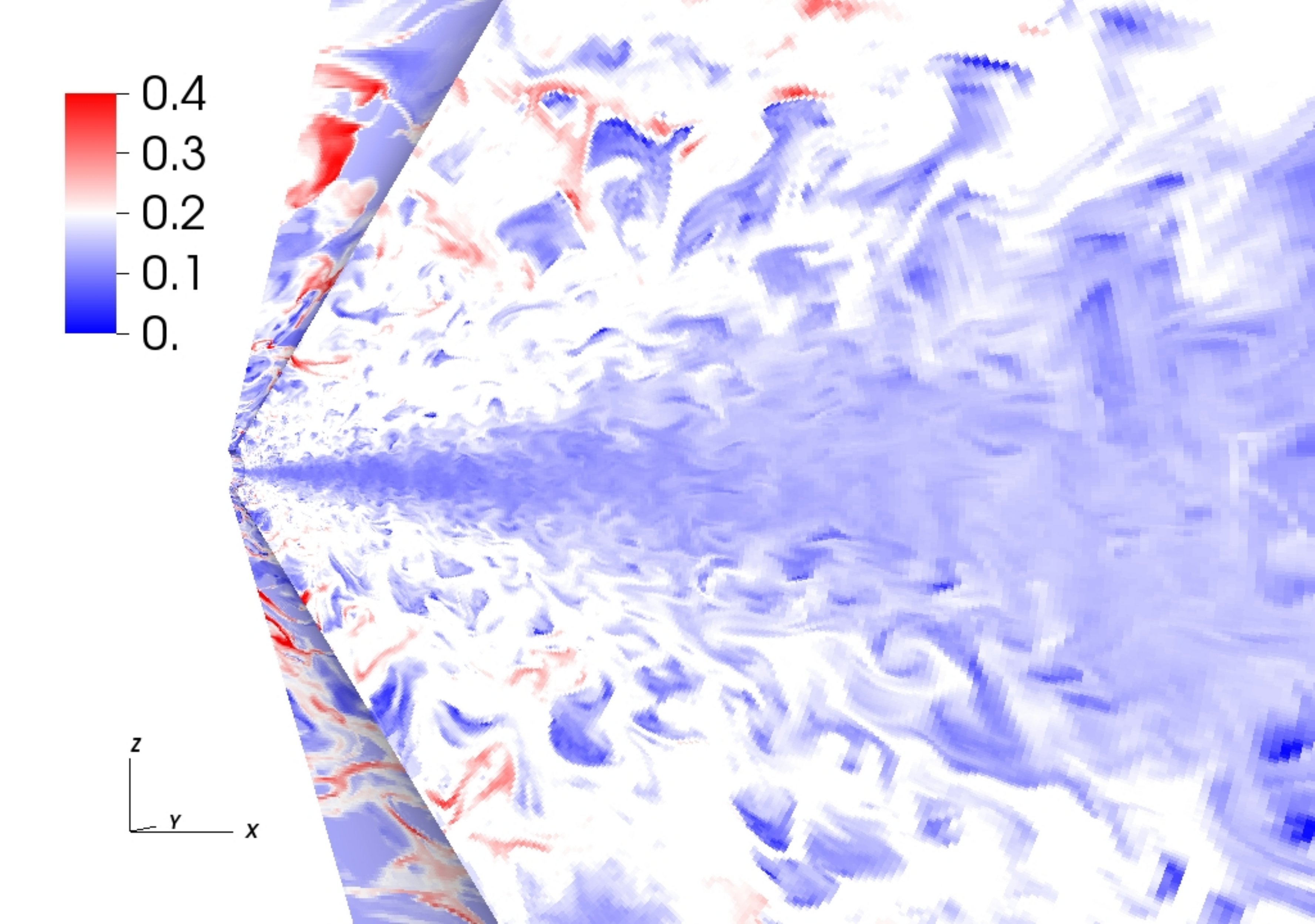}}
  \subfigure[$\rho$ at $t=3.98\times10^{5}\:GM/c^3$]{\includegraphics[width=0.5\textwidth]{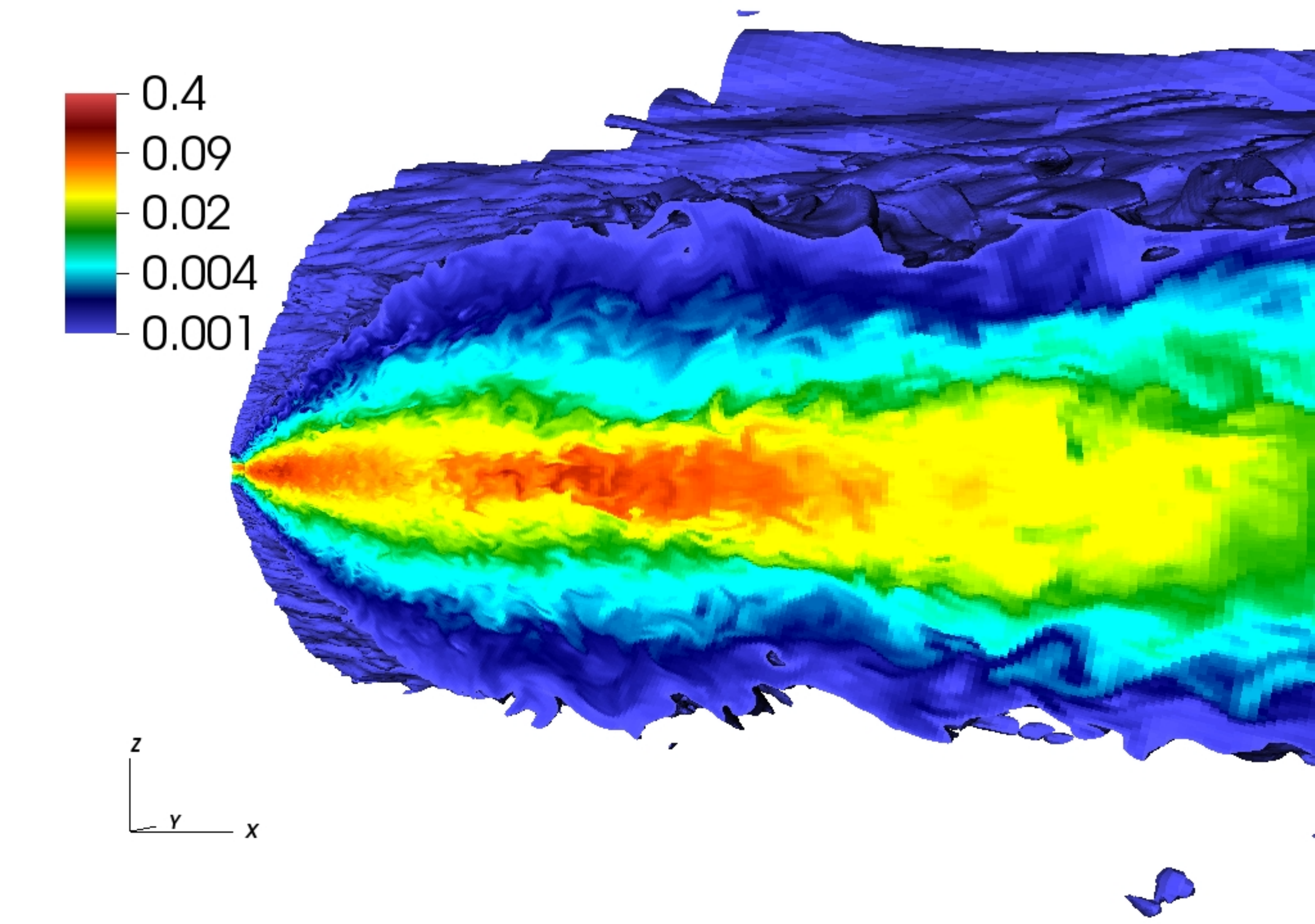}}
  \subfigure[$h_T/r$ at $t=3.98\times10^{5}\:GM/c^3$]{\includegraphics[width=0.5\textwidth]{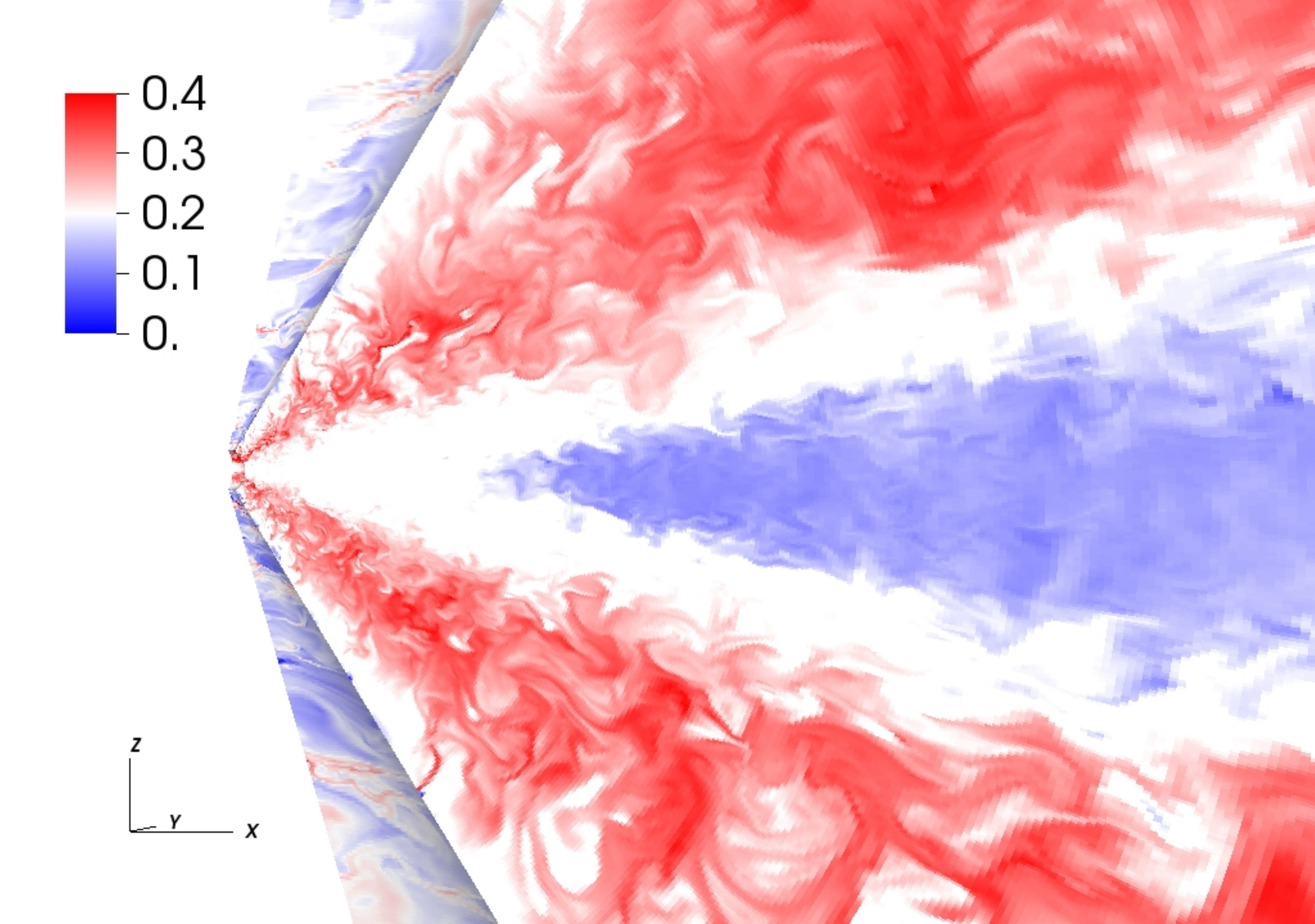}}
  \subfigure[$\langle \rho \rangle$]{\includegraphics[width=0.5\textwidth]{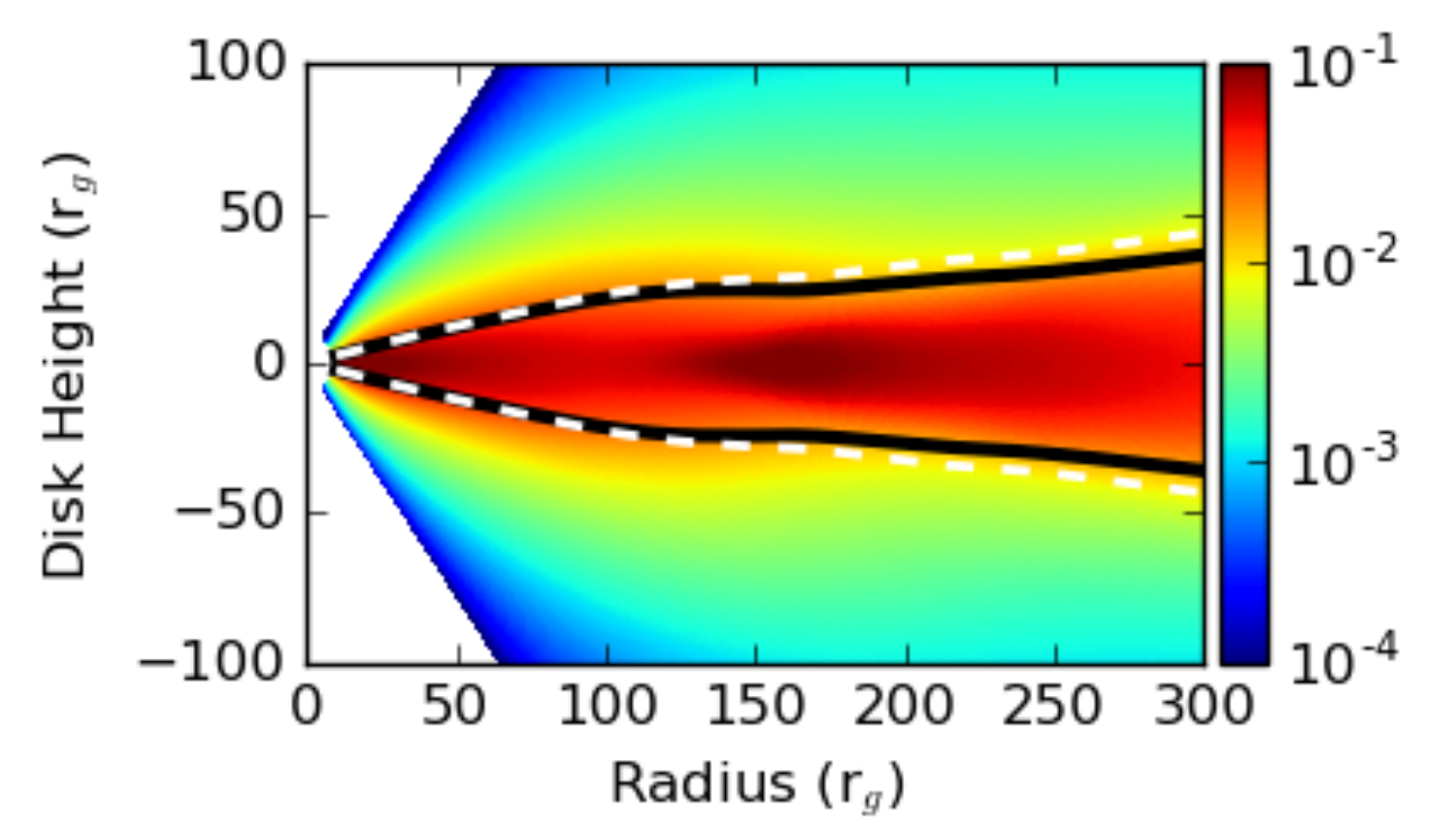}}
  \subfigure[$\langle h_T/r \rangle$]{\includegraphics[width=0.5\textwidth]{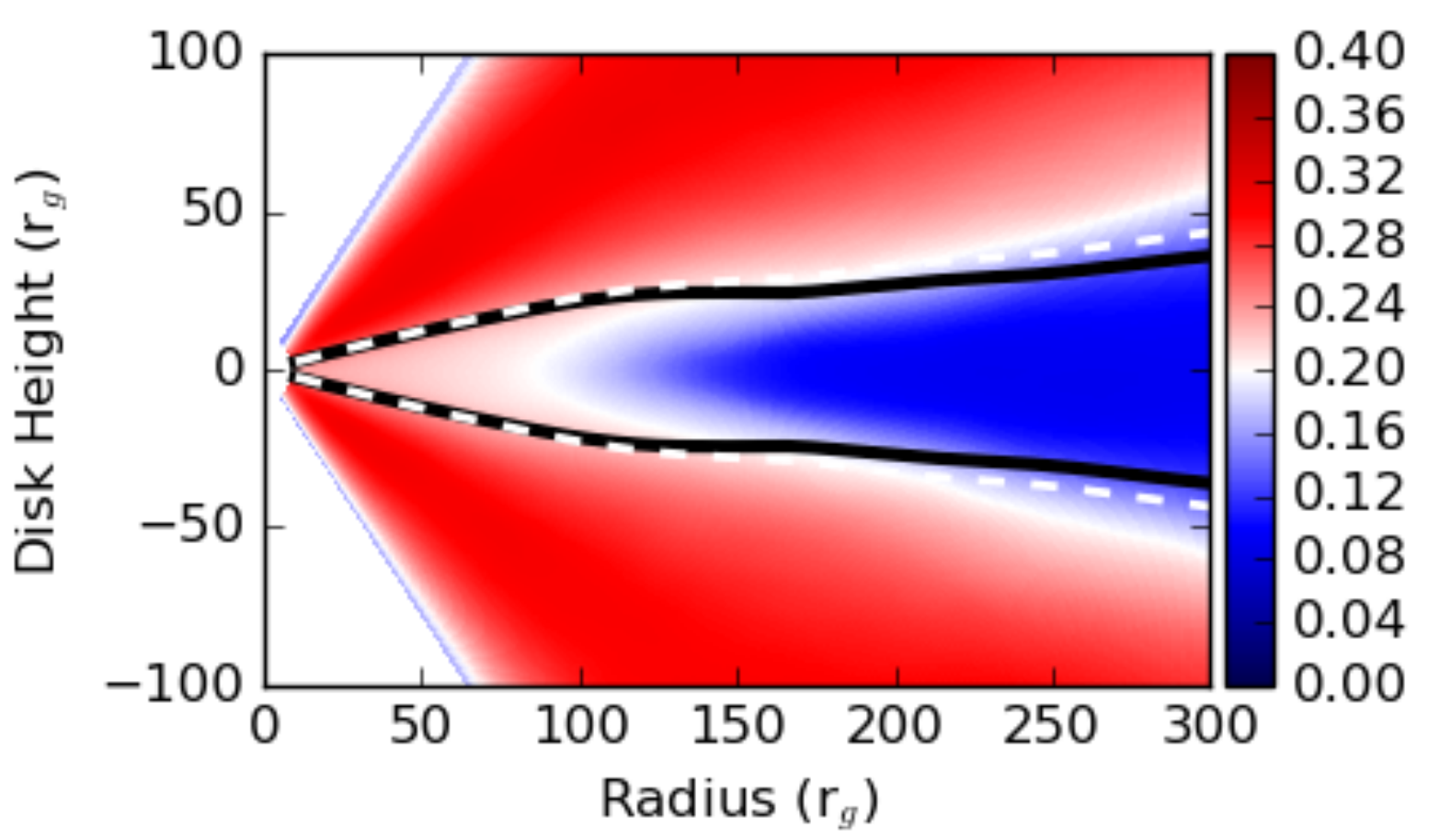}}
\caption{Top row- 3D renderings of density (left) and $h_{T}/r$ (right) at $t=1.99\times10^{5}\:GM/c^3$ (1000 $t_{orb,10}$).  The upper surface of the density rendering is set by the $\rho=0.001$ contour.  Middle row- 3D renderings of density (left) and $h_{T}/r$ (right) at $t=3.98\times10^{5}\:GM/c^3$ (2000 $t_{orb,10}$).  The upper surface of the density rendering is set by the $\rho=0.001$ contour.  Bottom row- Density (right) and $h_{T}/r$ (left) averaged over azimuth and time.  Both $\langle h_{g}/r \rangle$ (white dashed line) and $\langle h_{T}/r \rangle$ (black solid line )is overlaid in for reference.
\label{fig-trunc}}
\end{figure*}

We begin our analysis with an overview of the truncated disk in the simulation.  The truncated disk we study has hot, diffuse gas in the inner region and a cooler, Keplerian main disk body.  The inner region thickens since it is hot and the gas has additional thermal support.  Figure \ref{fig-scaleheights} shows two diagnostics of the time averaged scale height ratio.  The first measure is the geometric scale height,\begin{equation}
\frac{h_{g}(r)}{r} = \Bigg\langle \sqrt{\frac{\int (\theta(r) - \bar{\theta}(r))^2 \rho d\Omega}{\int \rho d\Omega}} \Bigg \rangle,
\end{equation} where $d\Omega=\sin \theta d\theta d\phi$ is the solid angle element in spherical coordinates and, \begin{equation}
\bar{\theta}(r) = \frac{\int \theta(r) \rho d\Omega}{\int \rho d\Omega}
\end{equation} is the average polar angle of the gas.  Second, is the density weighted, temperature scale height ratio,
\begin{equation}
\frac{h_{T}(r)}{r} = \Bigg\langle \frac{\int \sqrt{T r} \rho d\Omega}{\int \rho d\Omega} \Bigg\rangle
\end{equation} (where we have taken ${\cal R}=GM=1$). For each data dump, $h_{g}/r$ and $h_{T}/r$ were calculated and the time average was taken, denoted by the angled brackets.  At small radii, $r<50\:r_g$, $h_g/r$ and $h_{T}/r$ are $\approx0.25$.  This is below the target scale height of the upper branch of the cooling law ($h/r=0.4$) and similar to the peak of the average disk scale height in the HD model.  Like the HD model, advection acts as a non-radiative cooling term which helps to keep the disk thinner than the hot phase $h/r$ target.  At larger radii, the gas resides on the efficient cooling branch and maintains a thin geometry.

\begin{figure}
\includegraphics[width=0.5\textwidth]{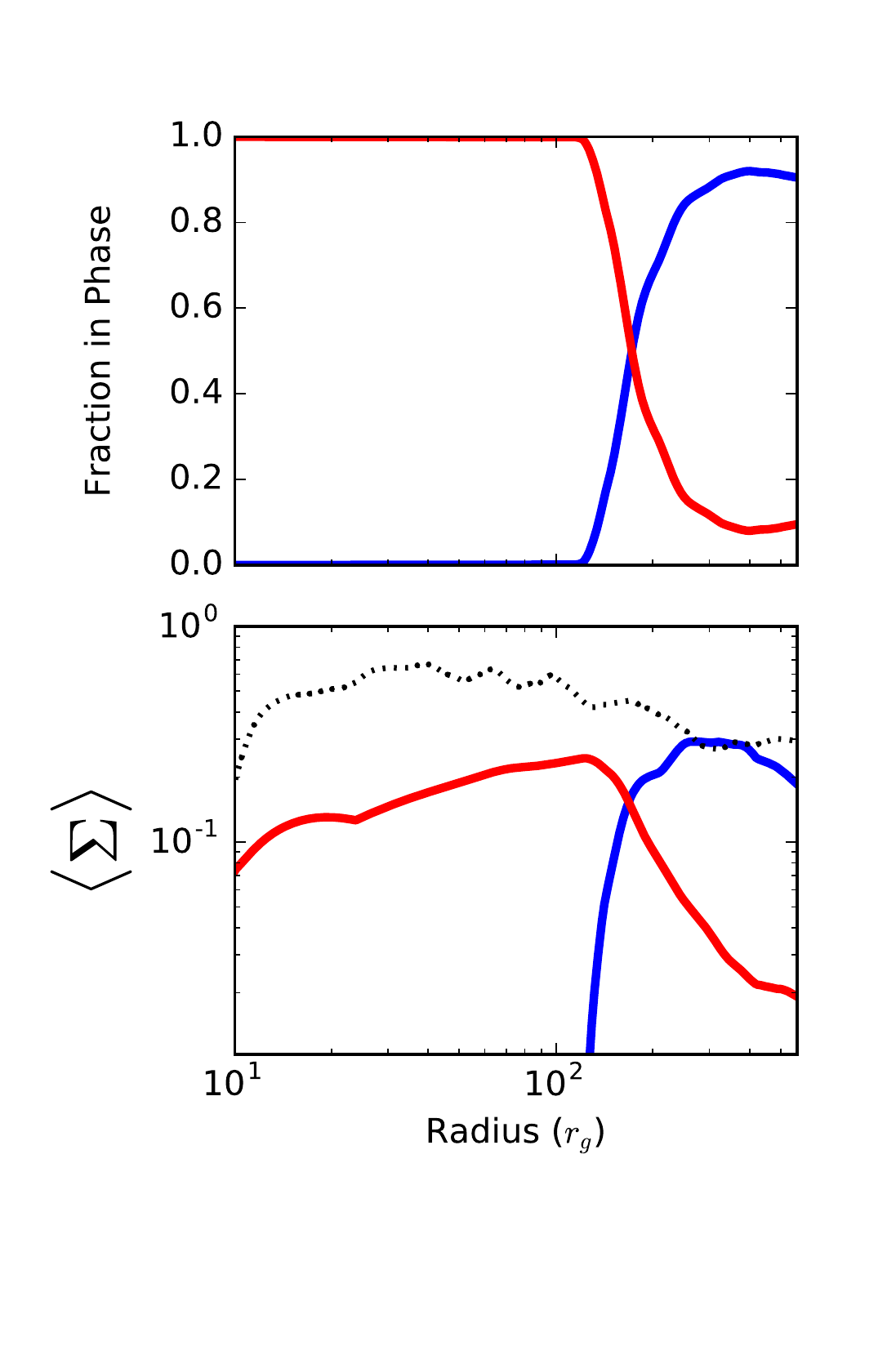}
\caption{Top panel- Radial profiles of the time-averaged $f_c$ (blue line) and $f_h$ (red line).  Bottom panel- Radial profiles of time-averaged $\Sigma_c$ (blue line) and $\Sigma_h$ (red line).  The surface density during the initialization immediately prior to the switch to the bistable cooling function is shown for reference as the black dotted line.  In the inner regions of the disk $\Sigma_c\approx 1\times10^{-7}$ and is off the plot axes.
\label{fig-gas_phase_and_den}}
\end{figure}

Figure \ref{fig-trunc} shows 3D volume renderings of $\rho$ and the local scale height temperature at $t=1.98\times10^{5}\:GM/c^3$ and $t=8.94\times10^{5}\:GM/c^3$, to highlight the changes the bistable cooling function introduces from the untruncated disk in the initialization.  Of the general differences, the decreased density and increased temperature of the gas in the inner disk are most evident.  While the inner region is evacuated, the density in the coronal regions increases.  Additionally, the gas in the atmosphere is significantly hotter and entirely in the hot phase, with temperature scale heights at or above $h_T/r=0.4$.  Throughout the simulation chaotic turbulence dominates the disk and at no point do coherent, small scale structures form.  Qualitatively, the truncation resembles the phenomenological model invoked to explain the spectral features from putative truncated accretion disks in BHB and AGN systems, e.g. the schematics in \citet{2004PThPS.155...99Z} and \citet{2007A&ARv..15....1D}, albeit with a hotter atmosphere.

Time and azimuthal averages of $\rho$ and the local $h_{T}/r$ are also shown in Figure \ref{fig-trunc}.  The gas density peaks in the disk midplane between $r=150-200\:r_g$.  In the truncated region the gas density is appreciably lower, but the disk is inflated so the density is enhanced vertically for a given radius.  Buoyancy introduces a stratified temperature structure in the outer disk.  The thickness of the cold, intermediate, and hot layers changes based on the radial heating profile.  The cold gas layer thins with decreasing radius as the gas interior to the truncation zone has $h_{T}/r \geq 0.2$.

Figure \ref{fig-gas_phase_and_den} shows profiles of the time averaged fraction of gas in each phase as a function of radius. The hot gas is taken to have $T>T_T$.  In the inner disk ($r<100\:r_g$), the gas is entirely above $T_T$.  In the outer disk ($r>300\:r_g$), $\approx 10\%$ of the gas resides in the hot phase.  The gas fractions gradually change in between these two regions as fluctuations allow varying amounts of the disk to reach $T_T$.  Quantitatively we take the ``truncation" to be at $r=170\: r_g$ where the two gas fractions are equal, but the region between $130-200\: r_g$ should really be considered a truncation zone between the outer, cold thin disk and the inner, hot thick disk.  This region corresponds to the tapering of the cold gas seen in the averaged $\rho$ and $h_{T}/r$ in Figure \ref{fig-trunc}.  The surface density,\begin{equation}
\Sigma(r)=\frac{1}{2 \pi r}\int \rho r^2 d\Omega,
\end{equation} of each respective gas phase is also shown in Figure \ref{fig-gas_phase_and_den}.  The integration has been done over the entire $\theta$ and $\phi$ domain at each radius.  In the inner disk $\Sigma$ is reduced by $70-80\%$ compared to the thin disk in the initialization stage.

\begin{figure*}
 \includegraphics[width=\textwidth]{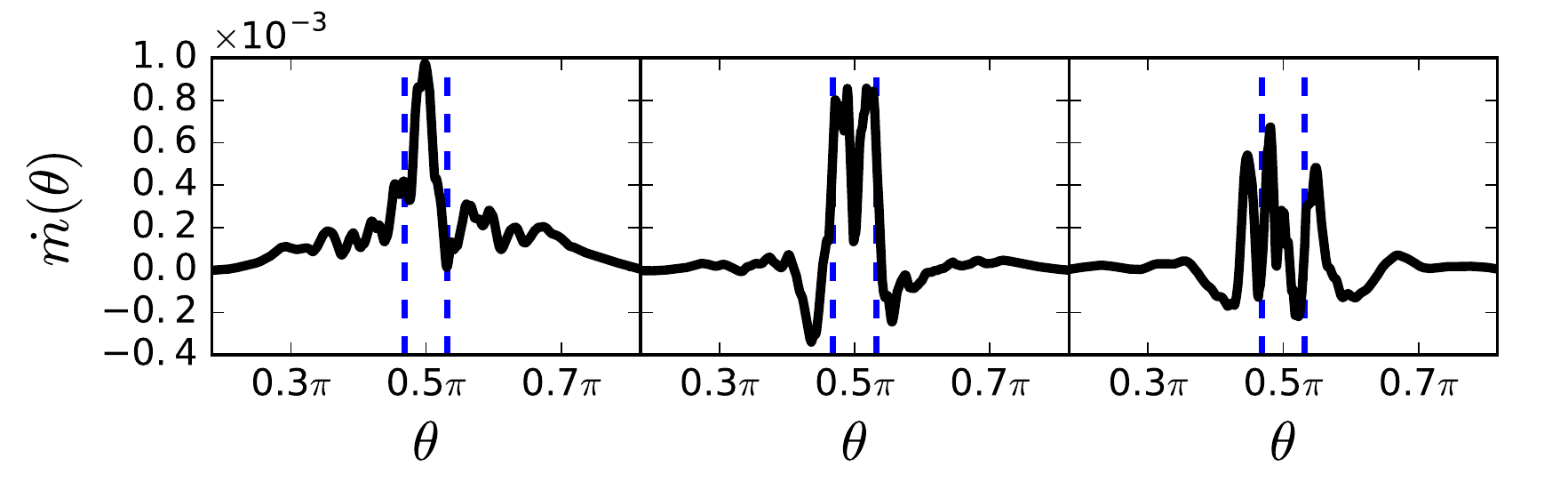}
\caption{Time averaged mass flux ($\dot{m}$) profiles along the polar angle ($\theta$) at $r=100\:r_g$ (left), $r=200\:r_g$ (middle), and $r=300\:r_g$ (right).  The blue dashed lines mark one disk scaleheight above and below the disk.  We adopt the convention that inflow is positive and outflow is negative.
\label{fig-wind_struct}}
\end{figure*}

\subsection{Disk Dynamics and Variability}

Three dynamically distinct regions develop in the simulation's global flow pattern.  The first is the thin, rotationally supported main disk body.  Radial motions are dominated by the chaotic flow, but are slightly negative on average as material slowly sinks deeper into the gravitational potential and accretes.  Inside of the transition zone, the flow behavior is slightly modified and $v_{\phi}$ is modestly sub-Keplerian (by $10\%$) because thermal support becomes more influential to the dynamics.  The radial velocity is greater than the slow diffusive drift found in the main disk body, but $v_r$ still remains less than a few fractions of a percent of the local Keplerian velocity.  In the hot gas filled disk atmosphere, however, the dynamics are quite different from the other two regions.  There, thermal support plays its biggest role and the gas is $10-15\%$ below the local Keplerian velocity.  The real distinguishing feature of this region, and most dynamically interesting, is a thermally driven outflow that underlies the turbulent fluctuations.

\begin{figure}
\includegraphics[width=0.5\textwidth]{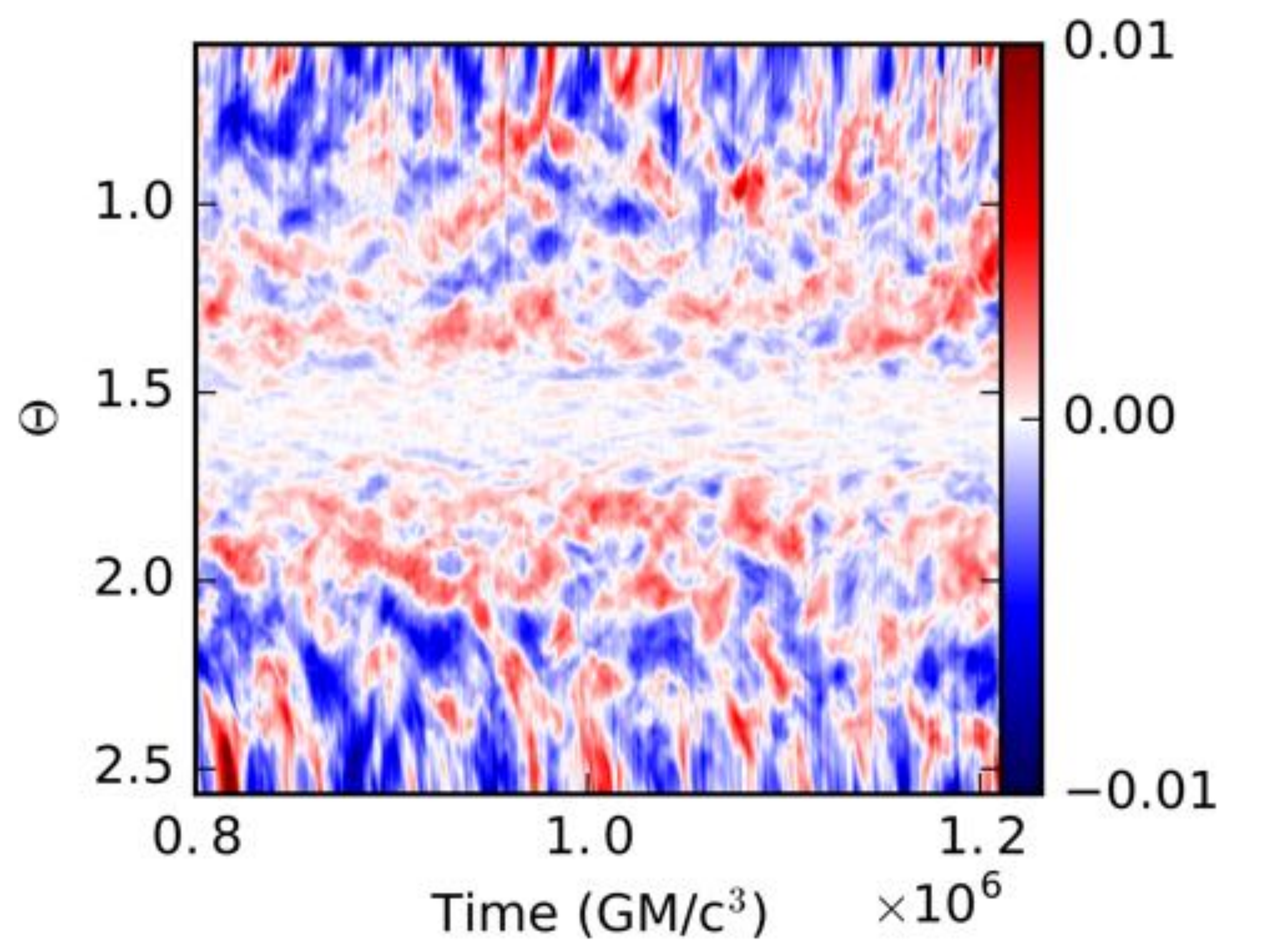}
\caption{Spacetime diagram of $v_r$ at $r=300\:r_g$.  Red denotes positive radial velocities and blue denotes negative radial velocities.  The disk midplane is found at $\theta=\pi / 2$.
\label{fig-outflow}}
\end{figure}

The vertical structure of the bulk gas dynamics is revealed through the mass flux as a function of polar angle \begin{equation}
\dot{m}(\theta)_r = -\int \rho v_{r} r \sin\theta d\phi, 
\end{equation} where $r$ is the radius of interest where the mass flux is calculated.  Figure \ref{fig-wind_struct} shows the averaged mass flux at $r=100, \:200, \:\&\: 400\:r_g$.  At $r=100\:r_g$, there is a time-averaged inflow as material and thermal energy is advected into the inner boundary, similar to an ADAF.  The profiles at $r=200\:\&\:400\:r_g$ show that this changes beyond the truncation.  Accretion still occurs in the main disk body where the mass is concentrated, but directly above and below the cold disk, starting at one disk scaleheight on either side, $\dot{m}$ is negative, indicating a net outflow of material.  This outflow is not a persistent, laminar wind.  Rather, it is a bulk motion in the highly turbulent gas from the thermal instability (compared to the strong wind in weak turbulence seen in \citet{2015ApJ...804..101Y}, for instance).  Figure \ref{fig-outflow} shows the spacetime diagram of the azimuthally averaged radial velocity at $r=300\:r_g$.  Consistent with the mass flux profiles, the outflow appears as the typically positive $v_r$ values directly above the main disk body, which is seen as the smaller amplitude fluctuations within $\theta=\pi/2\pm0.1$.  Moving away from the disk to higher altitudes, the fluctuations become more stochastic as the outflow strength diminishes.

\begin{figure}
  \subfigure[$\langle T(r) \rangle$ of Outflow]{\includegraphics[width=0.5\textwidth]{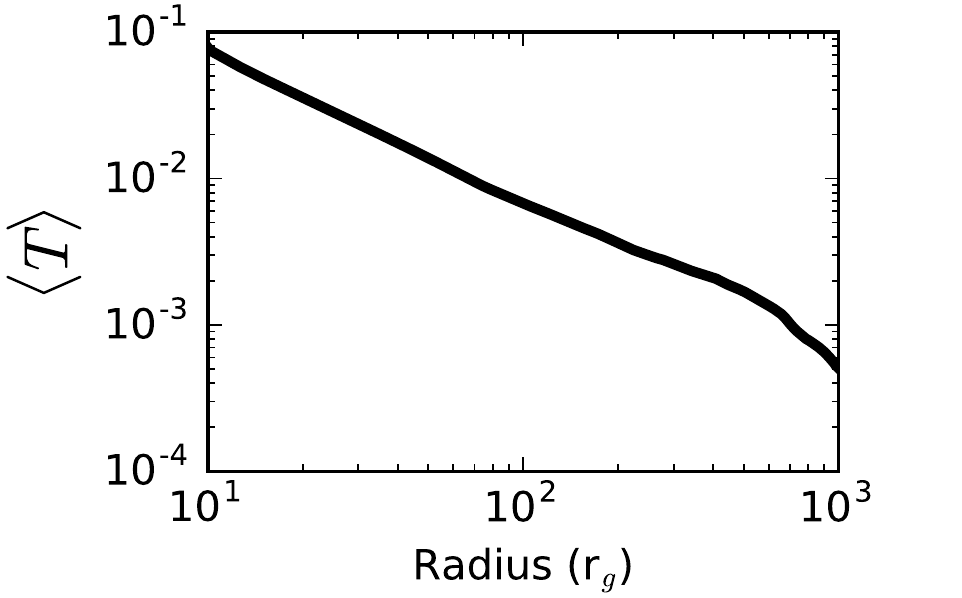}}
  \subfigure[$\langle v_r(r) \rangle$ of Outflow]{\includegraphics[width=0.5\textwidth]{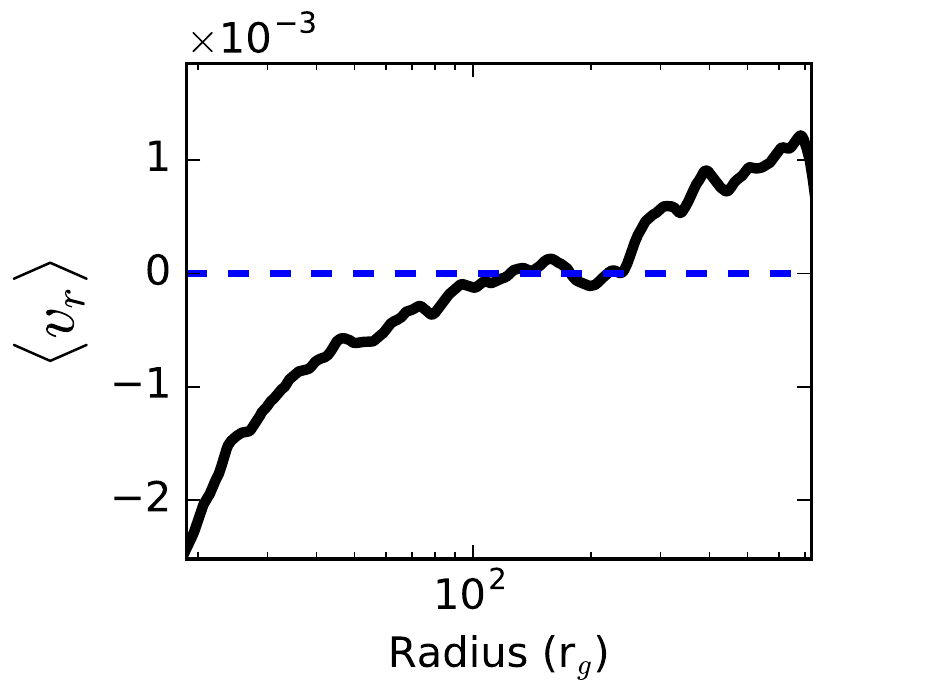}}
  \subfigure[$\langle \rho(r) \rangle$ of Outflow]{\includegraphics[width=0.5\textwidth]{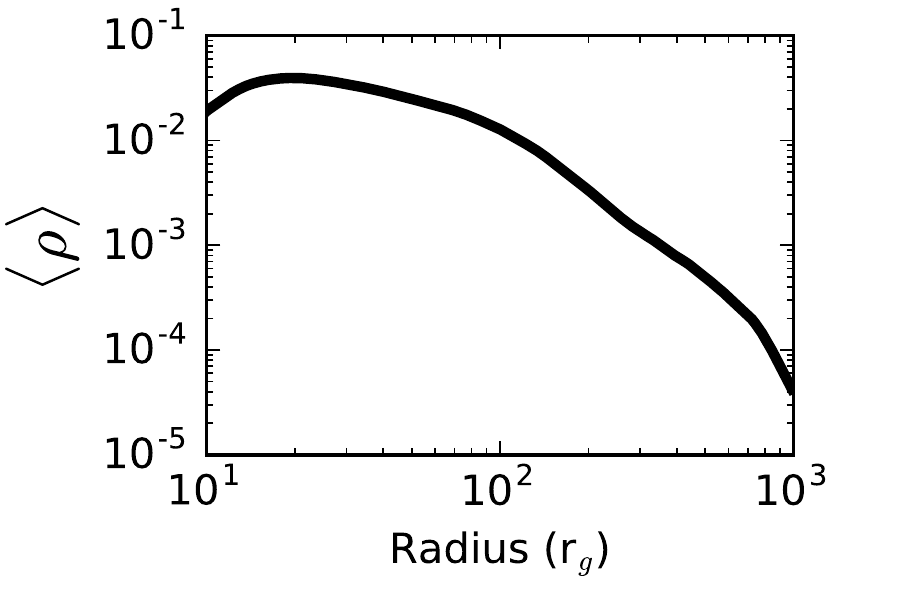}}
\caption{Radial profiles of the average gas temperature (a), radial velocity (b), and density (c) along the outflow ($\theta=\pi/2-0.3$ to $\theta=\pi/2-0.325$).  The profiles have been averaged over azimuth, height, and the duration of the simulation.  The blue-dashed line marks $v=0$, for reference.
\label{fig-wind_profiles}}
\end{figure}

The radial structure of the inflow in the inner disk and the outflow beyond the truncation zone is shown in the time averaged profiles of $T$, $v_r$, and $\rho$ in Figure \ref{fig-wind_profiles}.  We have taken a slice along the heart of the outflow from $\theta=\pi/2-0.3$ to $\theta=\pi/2-0.325$.  The temperature profile reveals a near perfect powerlaw scaling with index $\Gamma=-0.992\pm0.003$, consistent with the $T\propto r^{-1}$ solution proposed by \citet{1995ApJ...444..231N}.  The profiles of $v_r$ and $\rho$ display different behaviors on either side of transition zone.  In the inner disk the gas velocities are negative and accelerating inwards as material advects, and the density profile from $r=20-100 \: r_g$ is described by a $\Gamma=0.69\pm0.01$ powerlaw.  In the transition zone the average radial velocity is $v_r=0$ and the density profile rolls over.  Beyond the transition, the radial velocity of the outflow is positive and the gas feels a positive acceleration away from the black hole.  The acceleration of the outflow is due to buoyant forces and not the magnetic forces, which are quite small (a $10\%$ contribution).  The density profile steepens and falls off with a powerlaw index of $\Gamma=-2.13\pm0.01$.

\begin{figure}
\includegraphics[width=0.5\textwidth]{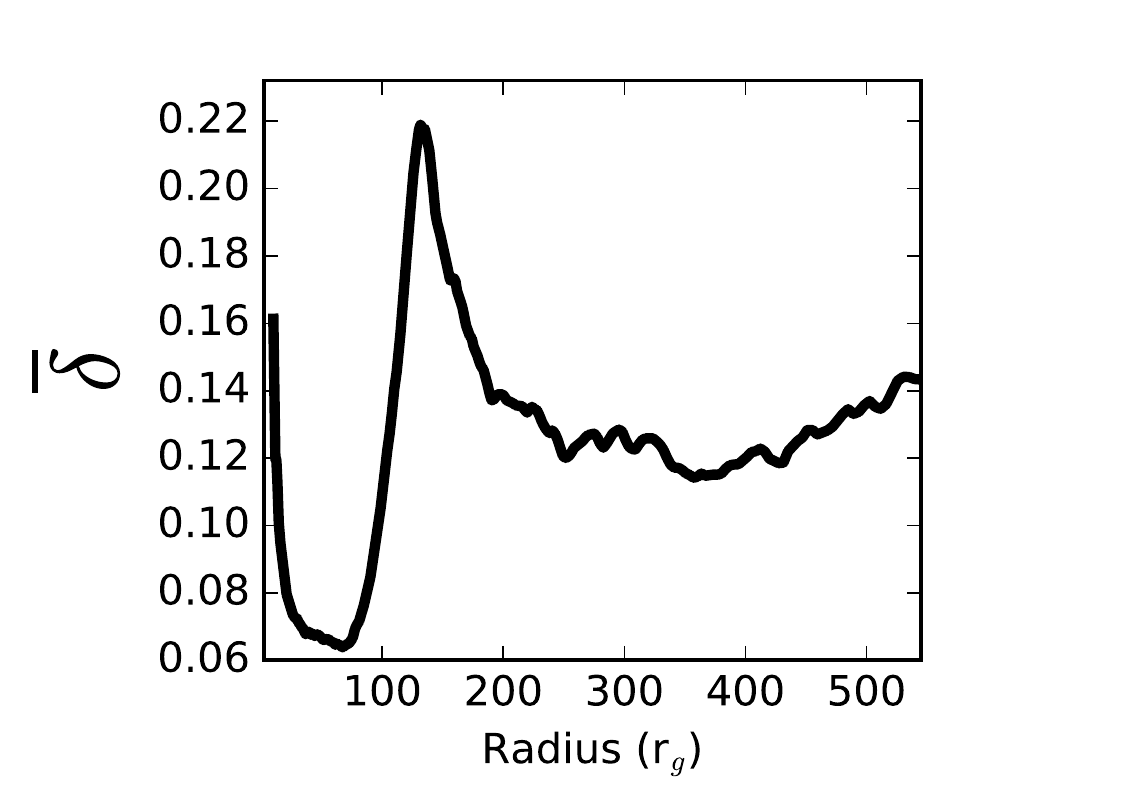}
\caption{Radial profile of the average turbulent intensity, $\overline{\delta(r)}$, measured in the density fluctuations in the $\phi$-direction at the disk midplane.  
\label{fig-rms}}
\end{figure}

The turbulent dynamics of the disk also display differences in the inner disk, transition zone, and outer disk.  A simple diagnostic of the turbulent density fluctuations is the turbulent intensity.  This is the time averaged, fractional standard deviation of the density as a function of radius, $\overline{\delta(r)}$, shown in Figure \ref{fig-rms}.  For each data dump, we calculate, \begin{equation}
\delta(r) = \frac{\sqrt{\langle | \rho_{mid}(r) - \langle \rho_{mid}(r) \rangle |^2\rangle}}{\langle \rho_{mid} \rangle},
\end{equation} with $\langle \cdot \rangle$ representing azimuthal averages and subscripts denoting that we only consider the disk midplane.  Considering the turbulent intensity offers several advantages for comparing the global behavior of the turbulence in the disk.  First, this essentially normalizes away any long-term variability in the density.  Second, turbulence is inherently a multiplicative process \citep{PhysRevE.58.4501,2007ApJ...658..423K} and, under constant conditions, the amplitude of fluctuations scales with the density so we need to remove effects from radial density gradients in the disk.

\begin{figure*}
\centering
\includegraphics[width=0.9\textwidth]{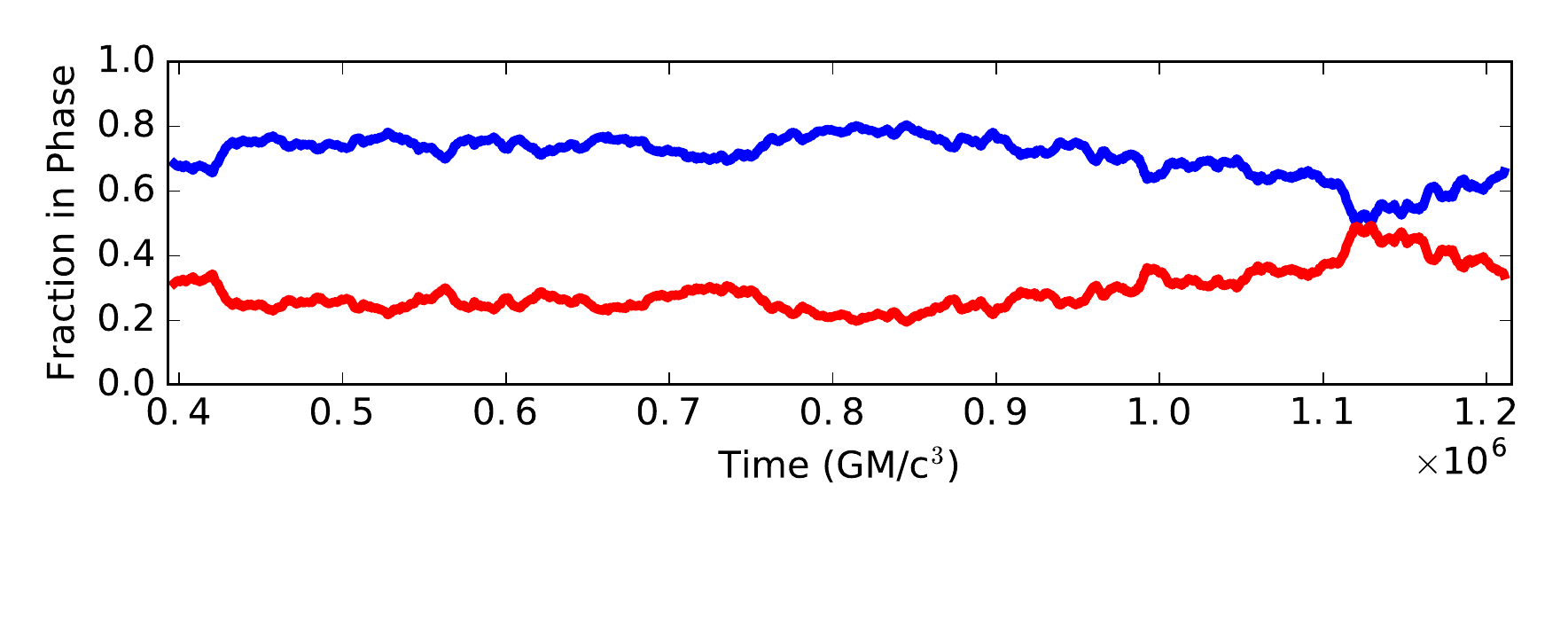}
\caption{Gas fractions $f_c$ (blue line) and $f_h$ (red line) at $r=200\:r_g$, just beyond the truncation.
\label{fig-frac_phase}}
\end{figure*}

Throughout most of the disk, $\overline{\delta}=0.13$, meaning that density fluctuations are on average $13\%$ of the mean.  As gas reaches the temperature transition in the truncation zone, $\overline{\delta}$ dramatically increases from $0.14$ at $r=200\:r_g$ to $0.22$ at $r=130\:r_g$.  There is then a precipitous drop in the inner disk and $\delta$ falls to $\overline{\delta}\approx0.06$ as the turbulence is suppressed.  Close to the inner boundary ($r<30\:r_g$), the turbulence increases again and approaches the intensity of the main disk body.

Searching for time variability is a goal of our analysis because it could offer a way to observationally probe the accretion flow and test the model predictions.  One key source of variability in a truncated accretion disk on timescales longer than the turbulent fluctuations is from appreciable changes in the gas fraction.  If this results from large changes in the thermal instability, it should appear on the local thermal time and potentially be observable.  Figure \ref{fig-frac_phase} shows the time evolution of the gas fraction in the hot phase, $f_h$, and the cold phase, $f_c$, at $r=200 \: r_g$.  Both $f_h$ and $f_c$ are featureless and meander around their average values of $0.29$ and $0.71$, respectively.  There is no obvious indication of an instability occurring on a thermal time in this simulation, where $t_{th}\approx2\times10^3\:GM/c^3$ at $r=200 \: r_g$.  The only prominent behavior is a gradual heating event in the latter half of the simulation.  At the peak of the heating event $f_h$ and $f_c$ are equivalent, but then the disk begins to cool and $f_h$ decreases.  Small amplitude, high frequency variability from turbulent fluctuations are superimposed on top of the long term trend.

\begin{figure*}
\centering
\includegraphics[width=\textwidth]{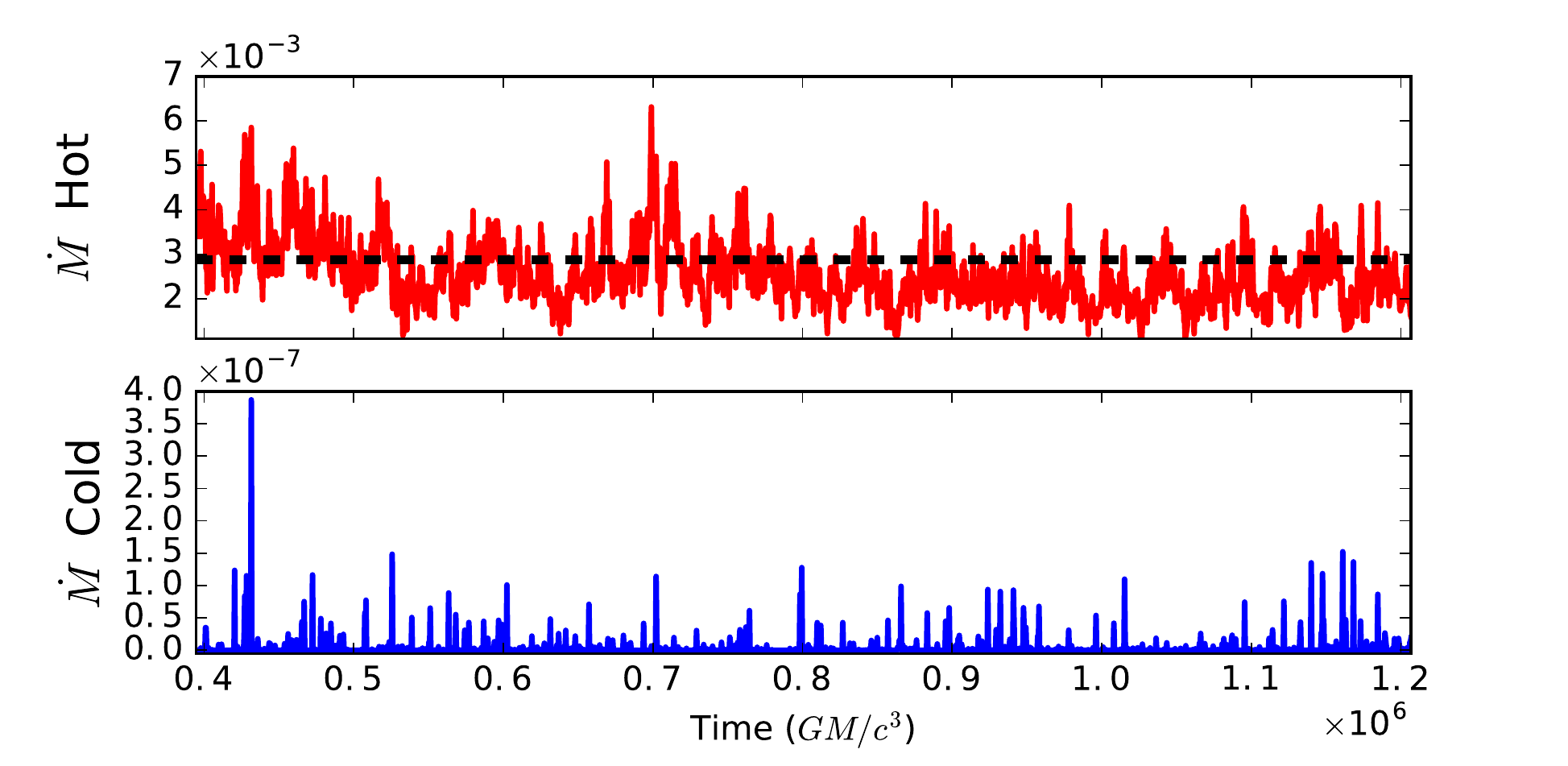}
\caption{$\dot{M}_h$ (top panel) and $\dot{M}_c$ (bottom panel) at the inner boundary over the course of the simulation.  The $\dot{M}_{init}$ at the inner boundary is shown as the black dashed line in the top panel for reference.
\label{fig-mdot}}
\end{figure*}

\begin{figure*}
\centering
\includegraphics[width=0.9\textwidth]{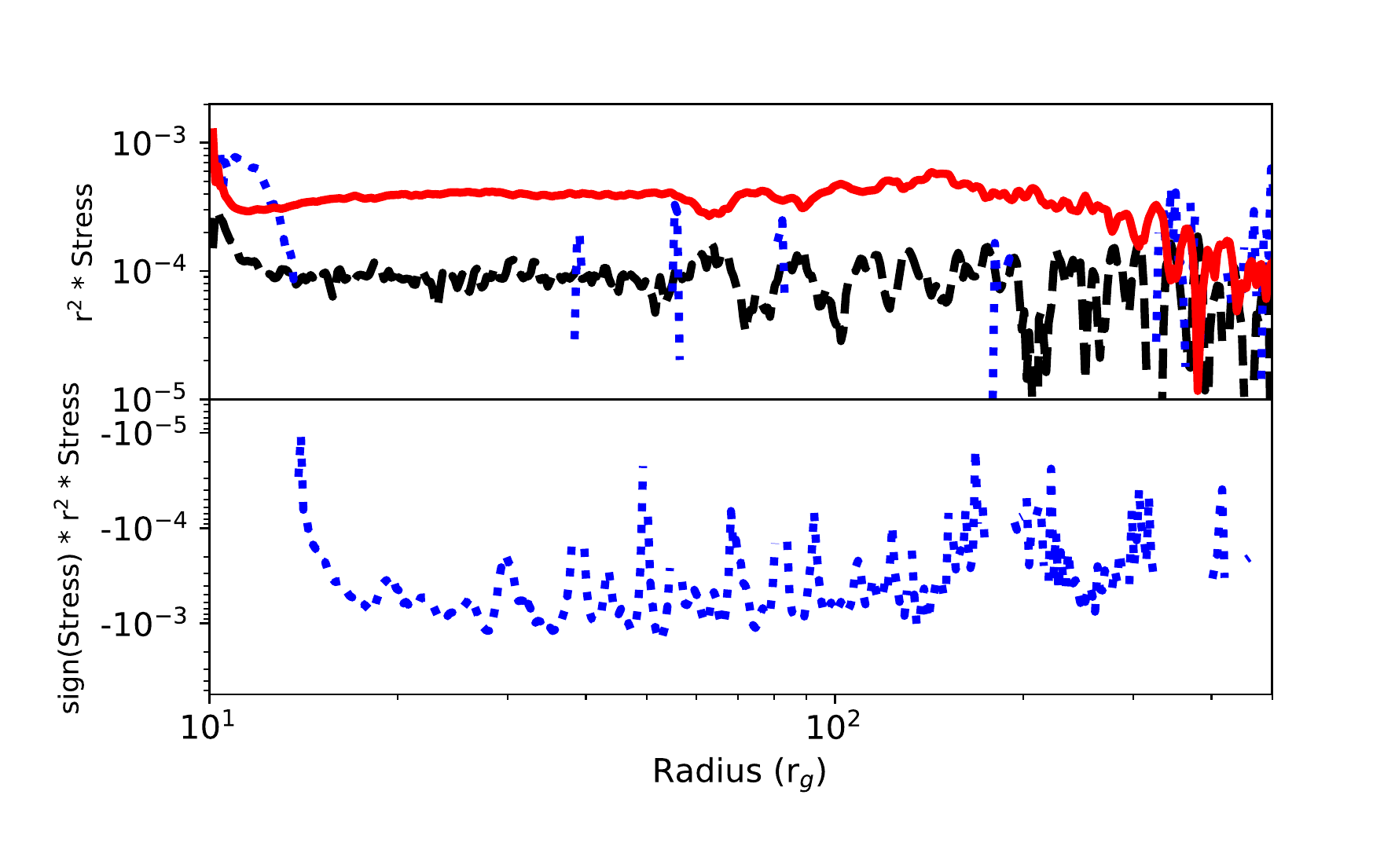}
\caption{Radial profiles of the shell integrated angular momentum advection (blue dotted line), Maxwell stress (solid red line), and the Reynolds stress (dashed black line).  The profiles are shown as the logarithm of the absolute value with the original sign of the stress, i.e. $\textrm{sign}(x) \textrm{log}(|x|)$.  Additionally, they have been scaled by $r^2$.
\label{fig-stress_profile}}
\end{figure*}

\begin{figure}[!htb]
\centering
\includegraphics[width=0.45\textwidth]{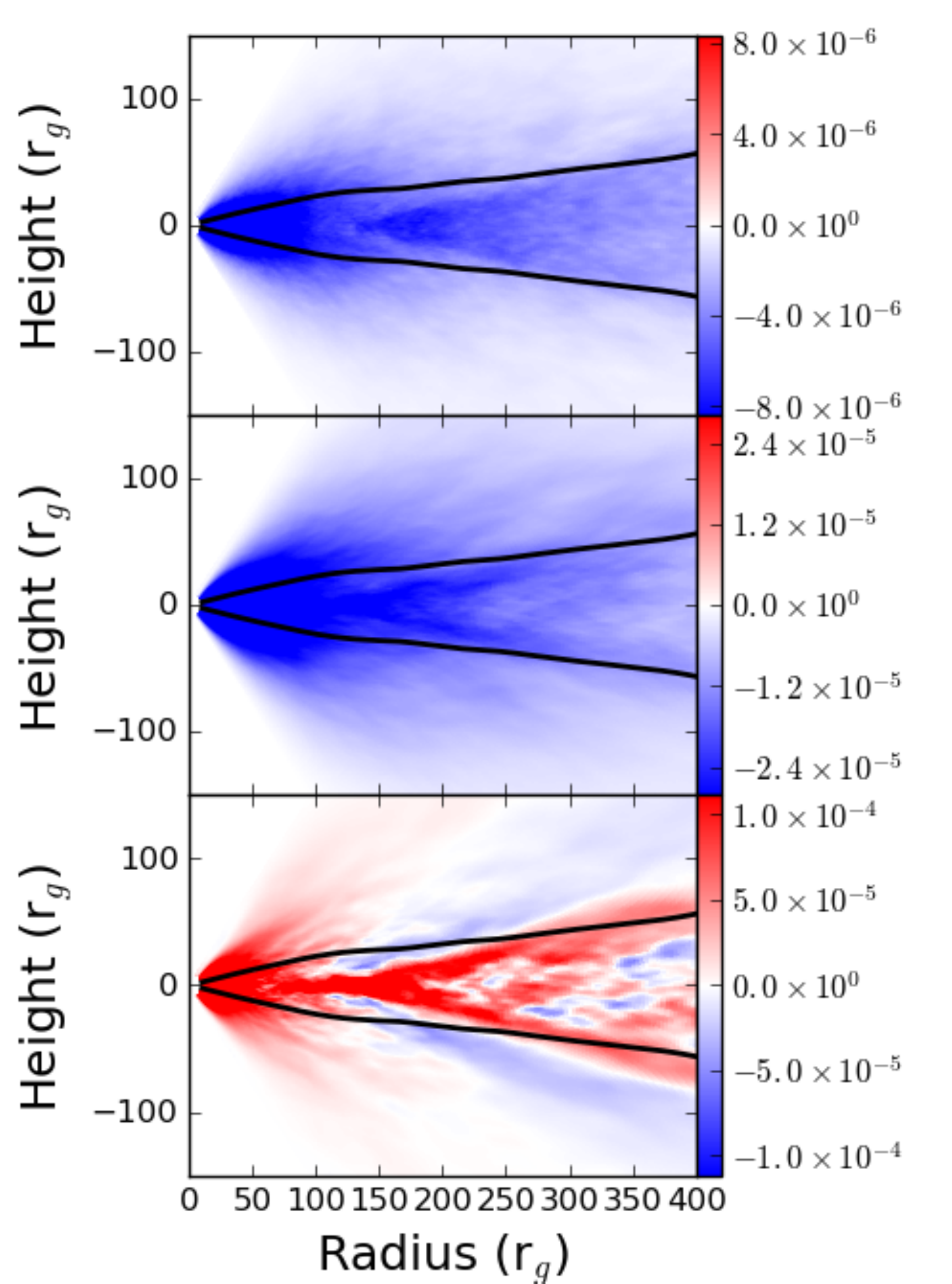}
\caption{Maps of $\rho r v_\phi \delta v_{r}$ (top panel), $-rB_r B_{\phi}$ (middle panel), and $\rho r v_\phi \langle v_{r} \rangle$ (bottom panel).  Overlaid is the average geometric scale height (black solid line).
\label{fig-stress_2D}}
\end{figure}

\subsection{Mass Accretion and Angular Momentum Transport}

The mass accretion rate, \begin{equation}
\dot{M}(r) = \int \rho v_{r} r^2 d\Omega, 
\end{equation} is highly variable on short timescales.  The variability in the simulation is from rapid fluctuations due to the disk turbulence, and there is a marked absence of any long-term, coherent variability that might be tied other accretion processes.  Figure \ref{fig-mdot} shows $\dot{M}$ at the inner boundary separated into the hot phase gas and cold phase components.  The hot gas is accreted at a rate of $\langle \dot{M_{h}} \rangle = 2.5\times10^{-3}$ with a characteristic variability of $\sigma_{\dot{M}}=7.0\times10^{-4}$, or $28\%$.  The cold gas accretion rate is negligible at $\langle \dot{M} \rangle=3.4\times10^{-9}$.  There are brief excursions from this very low value, but it rarely exceeds more than a $1\times10^{-7}$.  For reference, the mass accretion rate of the thin disk during initialization was $\langle \dot{M}_{init} \rangle=2.8\times10^{-3}$ is also shown.  The slight suppression in the truncated disk mass accretion is from the removal of a small amount of material by the weak outflow, which prevents it from reaching the inner boundary.

The transport of angular momentum enabling the mass accretion has three components: the Maxwell stress, the Reynolds stress, and advection.  Conservation of angular momentum is given by,
\begin{equation}\label{eqn-ang_mom}
\frac{\partial}{\partial t} ( \rho r v_\phi ) + \nabla \cdot (\rho r v_\phi \vvel -r B_{\phi} {\bf B}) = 0.
\end{equation} The dominant torque, $rB_r B_{\phi}$, is exerted by the Maxwell stress.  The $\rho r v_\phi \vvel$ term can be further separated into an average velocity term representing the advection of angular momentum, $\rho r v_\phi \langle v_{r} \rangle$, and its fluctuating component, $\rho r v_\phi \delta v_{r}$, which accounts for torques from the Reynolds stress.  The radial profiles of the three terms are shown in Figure \ref{fig-stress_profile} after summing over shells and taking the time average.

Throughout the disk the Maxwell and Reynolds stress transport angular momentum outwards to larger radii.  The Maxwell stress is $\approx5\times$ larger than the Reynolds stress and provides the majority of the angular momentum transport.  This is consistent with the 4-5:1 ratio broadly seen in other numerical models of MHD accretion disks \citep{1995ApJ...440..742H, 2006MNRAS.372..183P, 2008NewA...13..244B}.  Interior to the transition zone all of the angular momentum transport terms are enhanced by factors of 2 over what they are in the outer disk.

\begin{figure*}[!htb]
\centering
\includegraphics[width=0.85\textwidth]{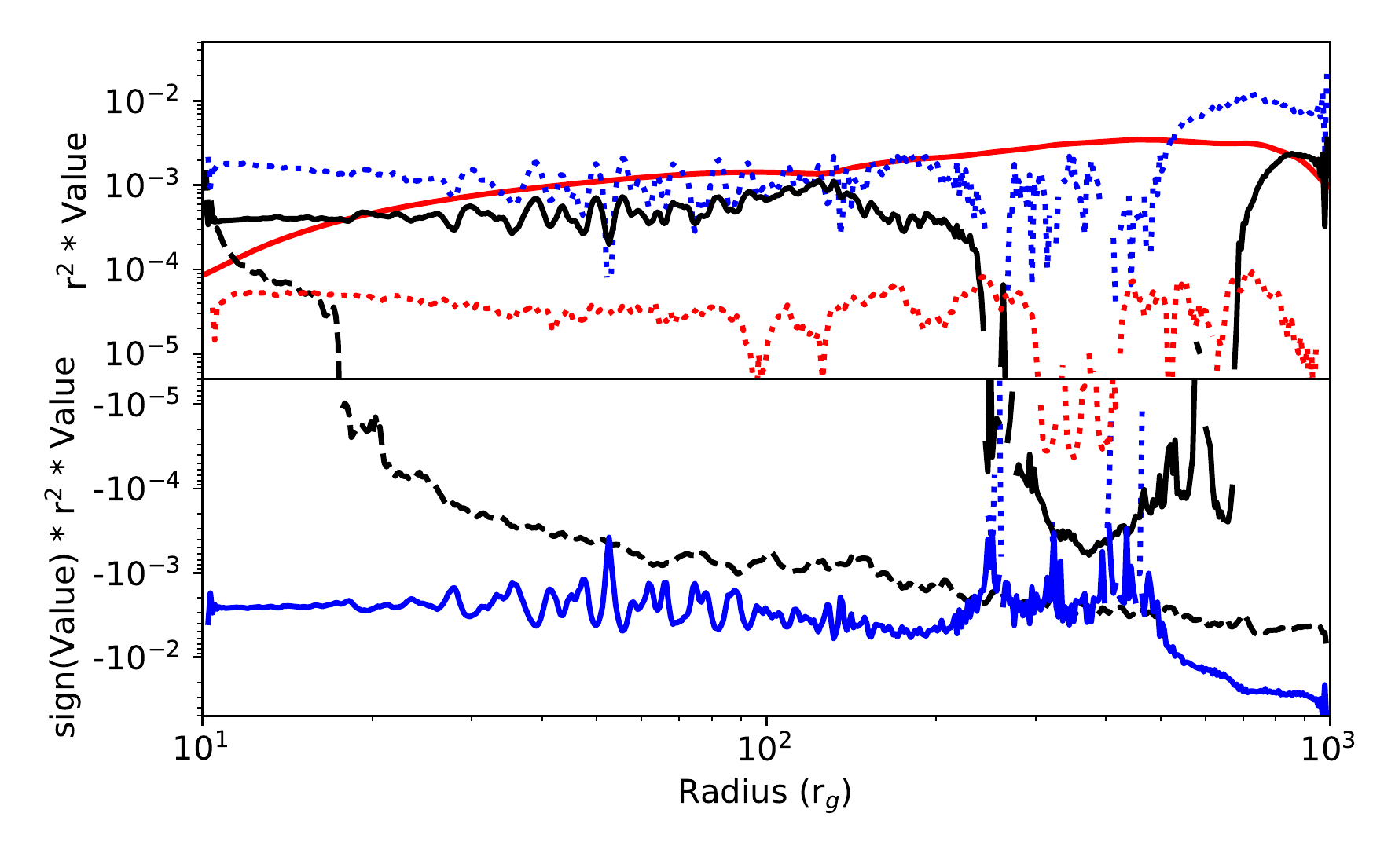}
\caption{Radial profiles of terms in the energy equation.  The profiles are shown as the logarithm of the absolute value with the original sign of the energy, i.e. $\textrm{sign}(x) \textrm{log}(|x|)$.  Additionally, they have been scaled by $r^2$.  The work done by magnetic torques ($\mathcal{W}$) is shown by the black dashed line, the advection of kinetic energy ($\mathcal{A_K}$) is shown by the blue dotted line, the advection of thermal energy ($\mathcal{A_{T}}$) is shown by the black solid line, the advection of magnetic energy $\mathcal{A_{M}}$ is shown by the red dotted line, the cooling term ($\mathcal{C}$) is shown as the red solid line, and the change in gravitational potential energy ($\mathcal{G}$) is shown by the blue solid line.
\label{fig-energy_profile}}
\end{figure*}

Spatially, there is a partitioning of the Reynolds and Maxwell stress enhancements and the increased stresses occur in discrete zones.  Figure \ref{fig-stress_2D} shows the two-dimensional structure of the average Reynolds stress, Maxwell stress, and advection.  The Reynolds stress is more intense within the disk body where transitioning gas causes the disk to taper.  This region coincides with the larger turbulent intensity, $\delta$.  Together, this indicates that the turbulent fluctuations not only have a larger amplitude, but they are also correlated.  

The Maxwell stress enhancement forms an envelope around the truncation zone.  Given that it occurs where the outflow originates and flows along the disk, it seems the outflow amplifies the field.  The likely mechanism behind this amplification is the conversion of kinetic energy into magnetic energy by nonlocal energy transfers through shells in Fourier space.  For a wavenumber of the velocity field $k_v$, and wavenumber in the magnetic field, $k_m$ in a turbulent flow, the magnetic field will receive energy from the velocity field for all shells with $k_v<k_m$ \citep{1978mfge.book.....M, 2011AnRFM..43..377M}.  Comparing the small turbulent scales in this region to the outflowing gas scale, the flow meets this criterion.  As the gentle outflow moves through the turbulent accretion flow, field lines are stretched and the field is amplified.  This phenomenon has been invoked in a number of contexts, including geophysical dynamo action of the terrestrial magnetic field \citep{2016EP&S...68...78P}, but is less well studied in accretion flows.  \citet{2011A&A...528A..17L} showed that these non-local transfers in spectral space operate in MRI-driven turbulence, but there has yet be an investigation into the global energy transfer properties of an accretion disk.

Several distinct regions form in the advective term.  The greatest inwards advection of angular momentum occurs on the surface of the disk, coincident with the largest Maxwell stresses.  Disk material typically accretes along the skin of the disk and then in a channel like region in the disk midplane stemming from the truncation.  It is important to bear in mind that the appearance of this channel-like region is simply the average advection and that entire region is turbulent, despite it being suppressed relative to the rest of the disk.  The inwards drift of material is still very slow and the gas dynamics are dominated by the turbulent velocity.  Above this inflowing region, the outflowing gas is launched in the transition zone as hotter gas rises to larger radii.  Interior to this region ($r<100\:r_g$), all of the gas is in the hot phase and the Maxwell and Reynolds stresses are large which drives material inwards to the inner boundary.

\begin{figure*}
\centering
\includegraphics[width=\textwidth]{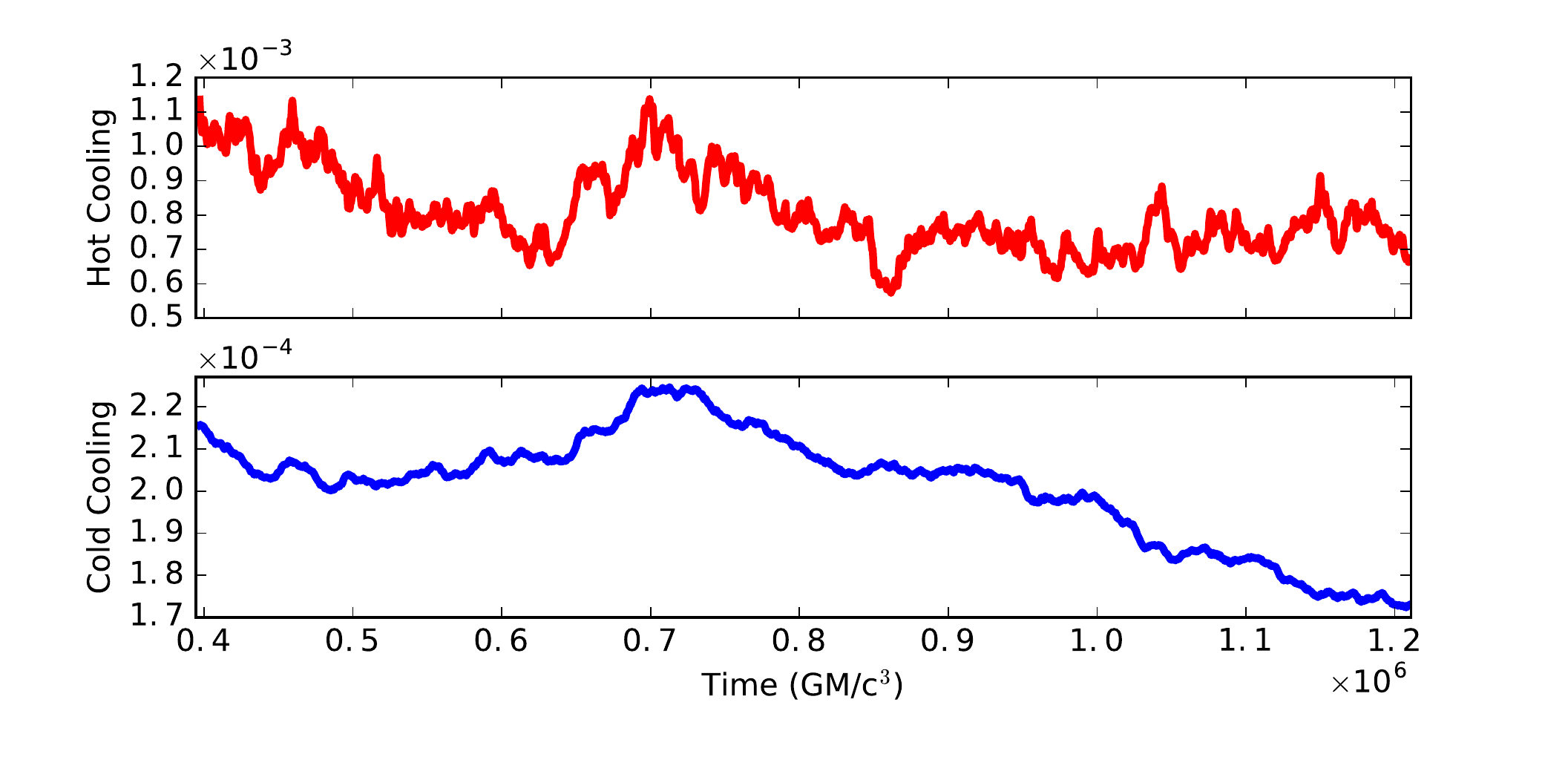}
\caption{Disk integrated synthetic light curve from the hot gas cooling (top panel) and cool gas cooling (bottom panel).
\label{fig-lightcurve}}
\end{figure*}

\subsection{Disk Energetics}

In this MHD accretion flow, the conservation of energy equation is, \begin{equation}
\frac{\partial \mathcal{E}}{\partial t}  + \mathcal{C} + \mathcal{A_{K}} + \mathcal{A_{T}} + \mathcal{A_{M}}+ \mathcal{G} - \mathcal{W} = 0,
\end{equation} where $\mathcal{E}$ is the total disk energy, $\mathcal{C}$ is the ``radiative" cooling done by the cooling function, $\mathcal{A_{K}}$ is the advection of kinetic energy, $\mathcal{A_{T}}$ is the advection of thermal energy, $\mathcal{A_{M}}$ is the advection of magnetic energy, $\mathcal{G}$ is the advection of gravitational potential energy, and $\mathcal{W}$ is the work done by magnetic torques.  Here, we will consider a spherical shell integrated form of the terms, such that they are  \begin{eqnarray}
&\frac{\partial \mathcal{E}}{\partial t}& = \frac{\partial}{\partial t} \bigg [ \int d\Omega \bigg(\frac{1}{2} \rho v^2 + \frac{P}{\gamma-1} + \rho \Phi + \frac{B^2}{2}\bigg)\bigg], \\
&\mathcal{C}& = \int d\Omega \Lambda, \\
&\mathcal{A_{K}}& = \frac{1}{r^2} \frac{\partial}{\partial r} \bigg[ r^2 \int d\Omega \bigg( \frac{1}{2} \rho v^2 \bigg)v_{r} \bigg], \\
&\mathcal{A_{T}}& = \frac{1}{r^2} \frac{\partial}{\partial r} \bigg[ r^2 \int d\Omega \bigg( \frac{P}{\gamma-1}\bigg)v_{r} \bigg], \\
&\mathcal{A_{M}}& = \frac{1}{r^2} \frac{\partial}{\partial r} \bigg[ r^2 \int d\Omega \bigg( \frac{B^2}{2}\bigg)v_{r} \bigg], \\
&\mathcal{G}& = \frac{1}{r^2} \frac{\partial}{\partial r} \bigg[ r^2 \int d\Omega \bigg( \rho \Phi \bigg)v_{r} \bigg], \\
&\mathcal{W}& = \frac{1}{r^2} \frac{\partial}{\partial r} \bigg[ r^2 \int d\Omega  v_{\phi} B_{\phi}B_{r}\bigg].
\end{eqnarray}  The time averaged radial profiles of $\mathcal{C}$, $\mathcal{A_{K}}$, $\mathcal{A_{T}}$, $\mathcal{A_{M}}$, $\mathcal{G}$, and $\mathcal{W}$ are shown in Figure \ref{fig-energy_profile}. 

In the inner disk the kinetic, thermal, and magnetic energy fluxes are negative, indicating there is an inwards advection with the accretion flow.  Most of the advection is in $\mathcal{A_K}$, followed by thermal energy ($\mathcal{A_{T}}\approx0.5\:\mathcal{A_{K}}$) and finally the magnetic energy ($\mathcal{A_{M}}\approx0.03\:\mathcal{A_K}$).  Beyond the truncation in the thin disk, $\mathcal{A_K}$ remains constant, but $\mathcal{A_{T}}$ actually switches sign as the hot gas in the outflow dominates this energy term.  $\mathcal{A_{T}}$ is slightly suppressed since there is competition between the inwards flow of cooler gas in the disk body and the hot gas with positive radial velocity, but ultimately the hot gas is main contributor.  Additionally, the $\mathcal{A_M}$ term switches sign because buoyancy lifts organized patches of overmagnetized gas as the disk dynamo cyclically builds magnetic field in the outer disk.  The role of magnetic buoyancy is minimal in the inner disk because the dynamo is suppressed, an aspect of the simulation discussed on Section \ref{sec-dynamo}. The contribution of $\mathcal{C}$ to the energy budget increases further out in the disk, reflecting the change in gas fraction and cooling efficiency.  In the inner disk, $\mathcal{C}\approx\:0.3\:\mathcal{G}$.  In the outer region, though, the disk approaches the standard thin disk ratio of $\mathcal{C}\approx1.5\:\mathcal{G}$.

Using the \emph{ad hoc} cooling as a proxy for emitted radiation, we create synthetic light curves, shown in Figure \ref{fig-lightcurve}.  The cooling is separated into that from the hot and cold gas phases, which can loosely be considered ``hard"-band and ``soft"-band emission, following observational convention.  The cool gas light curve has been renormalized to correct for the draining of cool gas from the simulation.  Over the course of the simulation, the cool gas depletes by $\approx20\%$.  We scale the density by the ratio of cool gas mass at the beginning of our analysis to the current cool gas mass at each point.  This scaling was not applied to the hot gas because we have reached an approximate inflow equilibrium so the rate at which it is accreted is the same rate at which cold gas transitions.  This keeps the total hot gas mass roughly constant throughout the analysis portion of the simulation.

The variability in these light curves predominately occurs on long timescales, i.e. longer than a dynamical or thermal time.  The main feature is a brightening event that peaks in both light curves at $t=7\times10^5\:GM/c^3$.  This event increases the cool gas light curve by only $10\%$, but the hot gas light curve by $50\%$.  The variability of the gas fraction near the transition zone seen in Figure \ref{fig-frac_phase} does not translate to the light curve.  The amplitude of stochastic fluctuations is much greater in the hot gas light curve.  Several flaring events occur when the hot gas light curve rapidly increases by $30\%$ and then decays to the lower trend level.  These events occur randomly and do not have any distinguishing periodicity or even a characteristic shape.

\subsection{Disk Dynamo}
\label{sec-dynamo}

The importance of the magnetic dynamo action on the evolution of a black hole accretion disk cannot be overstated.  On small scales an effective MHD dynamo is essential for sustaining the MRI driven turbulence because it provides a continual source of the magnetic field regeneration against dissipation and reconnection \citep[first shown by][]{1995ApJ...440..742H}.  On larger scales, spacetime diagrams of $B_{\phi}$ in numerical simulations reveal a quasiperiodic reversal of the magnetic field orientation and lifting of the field which forms the so-called ``butterfly" pattern.  Global PSDs show that, in fact, the organization of the field is a large-scale phenomenon and bands of power in the oscillation span several radii \citep{2011ApJ...736..107O}.  Nominally, the frequency of the oscillation, $\Omega_d$, is taken to be one tenth of $\Omega_K$.

\begin{figure*}
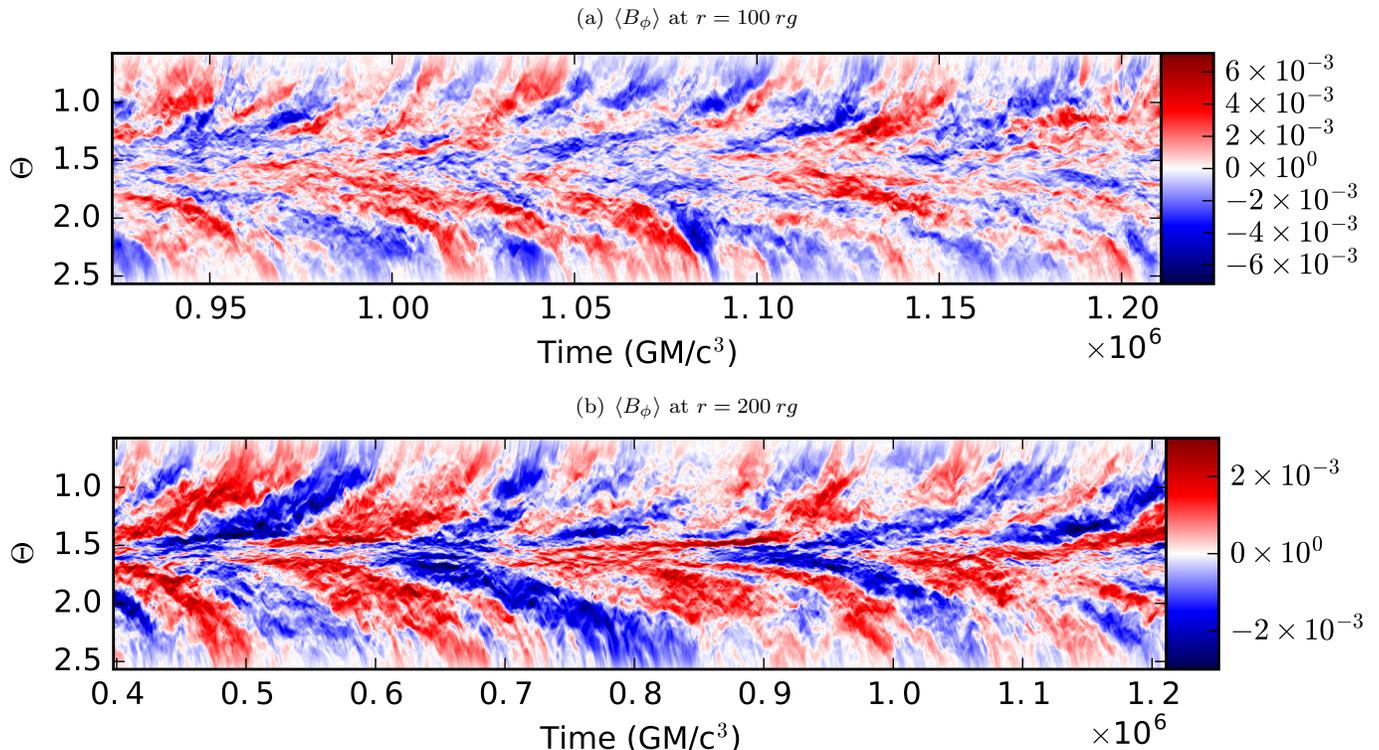

\centering
  \subfigure[$\langle B_{\phi} \rangle$ at $r=100\:rg$]{\includegraphics[width=\textwidth]{f14a.pdf}}
  \subfigure[$\langle B_{\phi} \rangle$ at $r=200\:rg$]{\includegraphics[width=\textwidth]{f14b.pdf}}
  \vspace{-0.5cm}
\caption{Spacetime diagrams of the azimuthally averaged $B_{\phi}$ at $r=100\:r_g$ (top panel) and $r=200\:r_g$ (bottom panel). The time axis in the top panel has been rescaled to show $r_{100}^{1.5}/r_{200}^{1.5}$ of the time to better match the differences in the evolutionary timescales of the dynamo process and aid in the direct comparison of $B_{\phi}$ at each radius.  These two disk locations were chosen to represent the more general behavior on either side of the truncation zone.
\label{fig-butterfly}}
\end{figure*}

Figure \ref{fig-butterfly} shows the spacetime diagram of the azimuthally average magnetic field on either side of the truncation zone at $r=100\:r_g$ and $r=200\:r_g$.  The time axis of the $r=100\:r_g$ butterfly diagram has been rescaled to show a similar evolutionary time (i.e. only a portion corresponding to $r_{100}^{1.5}/r_{200}^{1.5}$ of the analysis time).  Beyond the transition zone, the $B_\phi$ pattern is organized with a coherent, quasi periodic oscillation at a frequency of $f=3.7\times10^{-6}\:c^3/GM$ ($15\times$ the orbital frequency).  At the midplane the field strengthens, diminishes, alternates sign, strengthens and the cycle repeats.  The region is dominated by field of similar polarity spanning the midplane.  A cycle is present inside the truncation, but the pattern is disturbed and stochastic fluctuations account for much of the variability.  Additionally, the magnetic field is less ordered and the field oscillations are intermittent, thus preventing the development of strong field throughout the entire midplane.  Rather than sustained values that are typically near the maximum amplitude, weak $B_{\phi}$ is common.  Nevertheless, this does not lower the average Maxwell stress which is actually higher within the truncation zone, as we saw in Figure \ref{fig-stress_profile}.  At higher altitudes above and below the disk the cycle recovers and is more clearly defined.

\begin{figure*}
  \subfigure[$r-\theta$ map of $\langle \beta \rangle$]{\includegraphics[width=0.5\textwidth]{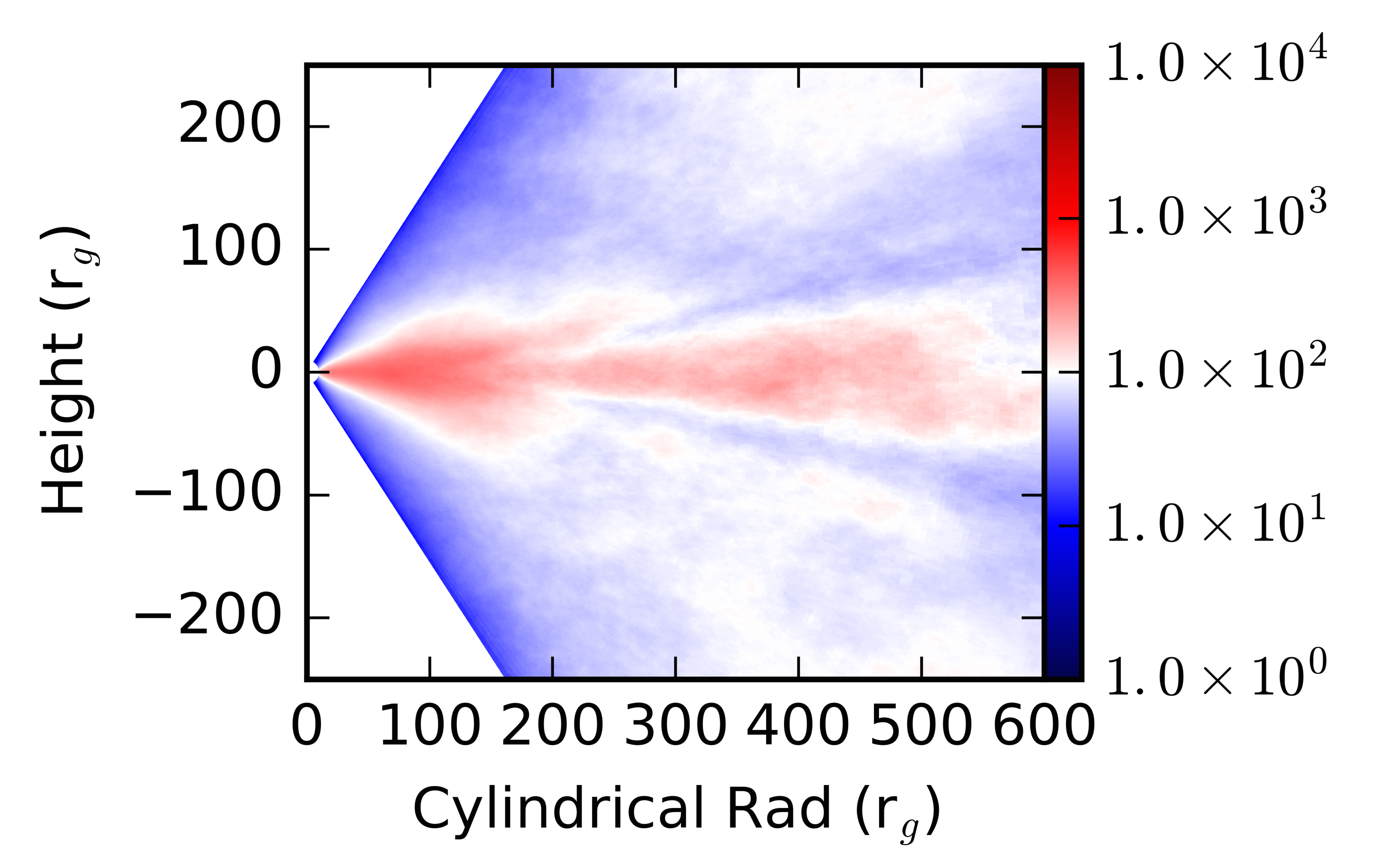}}
  \subfigure[Radial $\langle \alpha \rangle$ profile]{\includegraphics[width=0.5\textwidth]{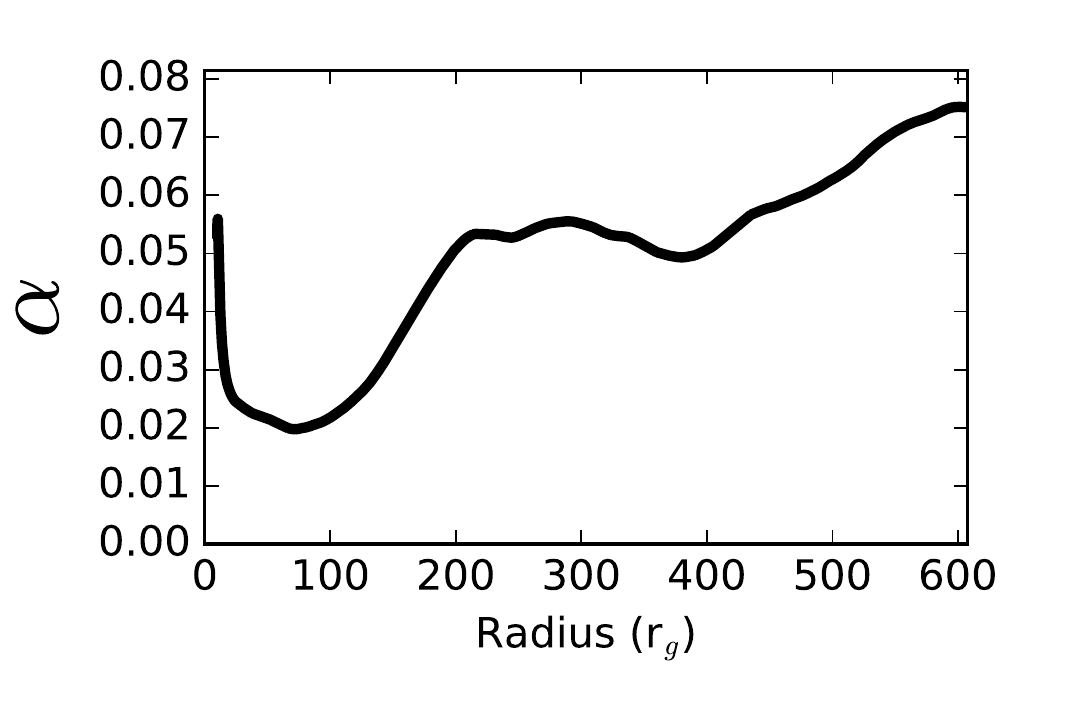}}
  \vspace{-0.7cm}
\caption{$r-\theta$ map of $\langle \beta \rangle$ (a) and the radial $\langle \alpha \rangle$ profile (b).  The $r-\theta$ map of $\langle \beta \rangle$ has been averaged over azimuth and time.  The radial $\langle \alpha \rangle$ profile is the time averaged effective $\alpha$ value calculated within the local geometric scale height ($h_g$).
\label{fig-beta_alpha}}
\end{figure*}

A consequence of disrupting the dynamo is that the net field strength is lower than it would be otherwise.  Coupled with the increased gas pressure from the transitioning gas, the ratio of the gas pressure to the magnetic pressure, \begin{equation}
\beta=\frac{P_{g}}{P_{m}},
\end{equation} is high in the inner disk and the transition zone.  Figure \ref{fig-beta_alpha} shows $\beta$ averaged over time and azimuth.  The increase in $\beta$ is most prominent in the inner region of the disk body where thermal pressure inflates the disk.  Extending away from the inner regions are wings of high-$\beta$ gas that trace the outflow.

The increased gas pressure also acts to decrease the effective $\alpha$-parameter in the inner disk.  Figure \ref{fig-beta_alpha} shows the time averaged radial profile of the effective $\alpha$.  For each data dump we calculated the volume averaged stress and volume averaged pressure within the geometric scale height ($h_g/r$) and took the ratio, according to Eqn \ref{eqn-alpha}. Beyond the transition in the region that has reached a quasi-steady state, the effective $\alpha$ parameter is $\alpha=0.055$.  This is $15\%$ below the average effective $\alpha$ in our calibration due to the additional gas pressure support from small patches of transitioning gas triggered by turbulent fluctuations.  Moving inwards from the truncation zone, the effective $\alpha$-parameter decreases from $200\:r_g$ to $110\:r_g$ which corresponds spatially with the tapering of the cold disk body seen in Figure \ref{fig-trunc}.  In the very inner regions where the gas is entirely in the hot phase, the effective $\alpha$-parameter settles to a value of $\alpha\approx0.02$.  There is a rapid recovery in the effective $\alpha$ at the inner edge of the simulation which could be due to boundary effects and thus artificial.

\begin{figure*}
\includegraphics[width=\textwidth]{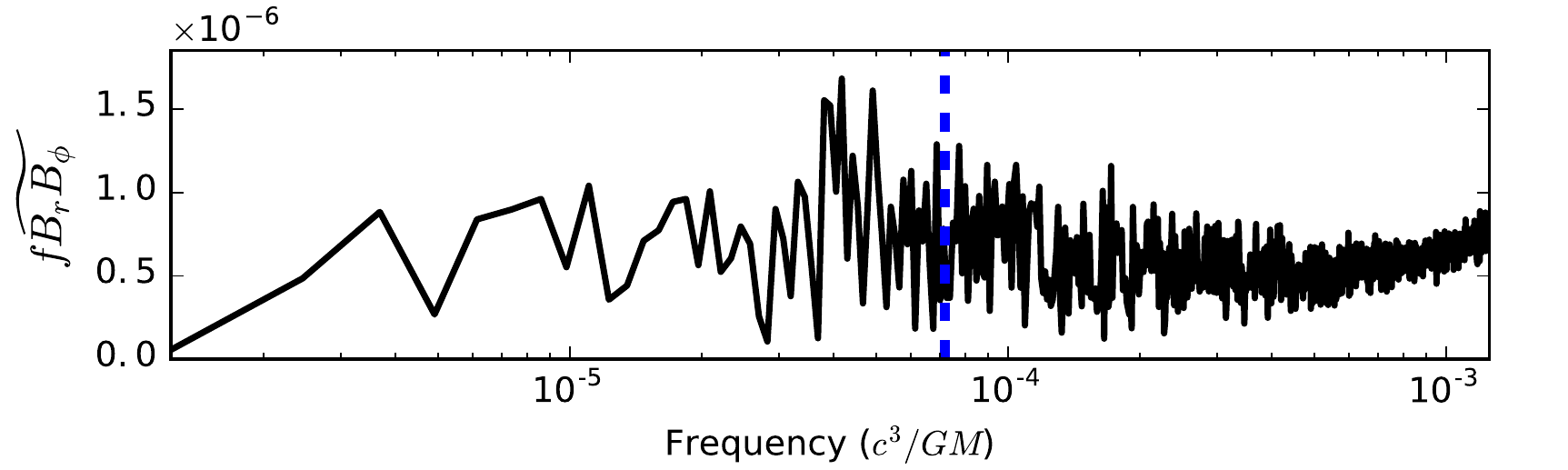}
\caption{Power spectrum of the Maxwell stress, $f\widetilde{M_{r \phi}}$, in the truncation zone.  The local dynamical frequency at $r=170\:r_g$ is shown as the dashed blue line.
\label{fig-brbphi_FT}}
\end{figure*}

The Fourier transform of the Maxwell stress,
\begin{equation}
\widetilde{M_{r \phi}}(f) = \int M_{r\phi}(t) e^{-2 \pi i f t} dt.
\end{equation} reveals periodicity in its variability.  Shown in Figure \ref{fig-brbphi_FT} is the power spectrum of the integrated Maxwell stress in the truncation zone between $r=130\:r_g$ and $r=200\:r_g$.  The total stress has been summed over $\theta={\pi\over2}\pm0.2$.  Overall, the power spectrum scales as $f^{-1}$ and $f\widetilde{M_{r \phi}(f)}$ is flat.  On top of the powerlaw spectrum, though, there is a band of power where the variability is strongly periodic.  This feature is centered at $f=4.0\times10^{-5}\:c^3/GM$, coherently spans five spectral bins, and is $\approx4\times$ greater than the noise spectrum.  Given its strength and relatively low frequency, this develops from an interaction with the disk dynamo.  Interestingly, it is five times faster than the local dynamo frequency at $f=7.2\times10^{-6}\:c^3/GM$ for $r=170\:r_g$.

\section{Discussion}
\label{sec-discussion}

The MHD model presented here and the viscous HD model in Paper \RNum{1} are targeted to study the dynamics of truncated accretion disks.  Compared to the rich literature on hot, geometrically thick accretion flows and cool, geometrically thin accretion flows, the truncated disk model is poorly explored, despite the frequency with which this flow configuration is invoked to explain the spectral properties of hard state BHBs and LLAGNs.  This work has characterized the general flow properties, but further work is required to fully capture the subtleties of flow at the transition zone.  Broadly, the HD and MHD models were similar even though they were set in very different fluid regimes.  Both models preferred a cool solution until near the imposed $r_T$, launched some sort of thermally driven outflow, and formed a thin layer of enhanced stress as the hot gas moved along the surface of the main disk body.  In the viscous HD model the enhanced stress was due to the $\theta-\phi$ viscosity component and in the MHD model this arose through the stretching of field lines which intensified the Maxwell stress.  These are two fundamentally distinct origins, but they ultimately have the same effect.  Furthermore, the inner disk regions in both models are evacuated, which is seen as deficits in the surface density profiles interior to $r_T$ compared to what it would be as a standard thin, cold disk.  This is generally consistent with observational results where the inner regions of hard state BHBs and LLAGNs are measured to be less dense. 

In the HD and MHD models the flow takes on a configuration where the cooler thin disk is embedded within the hot gas.  Typically, phenomenological models of truncated disks envision the hot gas occupying the inner region of the disk in a spherical bulge whose extent is roughly the truncation radius \citep[e.g.][]{1976ApJ...204..187S}.  Our results, along with other numerical models of truncated disks \citep[e.g. ][]{2016ApJ...826...23T, 2016MNRAS.463.3437M}, suggest that the hot gas may fill the entire region surrounding the black hole and take on a structure akin to that presented in \citet{1984SSRv...38..353L, 1997ApJ...489..865E, 1997ApJ...474L..57C}.  Additional scattering resulting from ``sandwiching" a disk between between the two hot coronal regions can produce telling observational signatures in the X-ray spectrum \citep {2016ApJ...826...23T} and in polarization \citep{2010ApJ...712..908S}.

While the presence of an outflow is shared between the HD and MHD truncated disk models, the manifestations are distinct.  In the HD model convection appears in the flow atmosphere, reminiscent of a CDAF-type flow, except that the gas in these convective eddies eventually fell back and was accreted.  In the MHD model convection is stabilized and, even though the instantaneous gas motions are dominated by MHD turbulence, there is a thermally driven bulk outflow.  Scaling the system to physical values, there appears to be limited promise in observing the outflow.  Taking a characteristic inner disk temperature to be virial at $T=10^{12}\:K$ \citep{1976ApJ...204..187S}, the outflow would have temperatures between $T\approx5\times10^9-10^{12}\:K$.  Scaling the dimensionless velocity by the speed of light, $c$, we find that the velocity range spans $v_r\approx0-300$ km/s.  The velocity is in the range of empirically measured outflows, but its observational prospects may be hindered by the high temperature and the large turbulent broadening since the turbulent velocity is three times greater than the bulk flow ($\Delta v \approx 1,000$ km/s).

Considering the truncated disk in MHD has elucidated several unique disk features that have not previously been considered.  First is the interplay between stochastic temperature fluctuations and the thermal instability.  MHD turbulence driven by the MRI has been understood as a key way to transport angular momentum and heat the disk through turbulent dissipation, and we find it is additionally important in triggering the thermal instability.  At a given radius, the gas has a range of temperatures with some fraction of the distribution lying beyond the transition condition.  This can impact the disk geometry because local pockets of gas can transition at radii which would otherwise be too cool to satisfy the thermodynamic criterion.  As gas sinks to smaller radii, the fraction of gas on the inefficient cooling branch will increase.  The natural result of this radial variation is a disk tapering in a transition zone like that produced in our simulation, rather than a transition edge.  Measurements of truncation radii from spectral fits may benefit from fitting a smoother transition rather than a sharp inner disk cutoff, as is often used.

The details of the transition rests on the precise transition mechanism.  Here, we treated the cooling with a simple threshold, but a more physical prescription is needed before a complete understanding of the dynamics in a truncated disk can be achieved.  This demands that the theory underlying the thermal instability be developed.  So far, several global accretion models have been proposed as candidates to fill this role.  Inner disk evaporation of  the cool, Keplerian disk by heating from a hot corona \citep{1994A&A...288..175M, 1996ApJ...457..821N, 1999ApJ...527L..17L, 2000A&A...360.1170R} could lead to disk truncation.  The interior of the disk could also heat the gas beyond the virial temperature because of the diffusion of energy from through the disk \citep{1996PASJ...48...77H, 2000ApJ...529..127M, 2000ApJ...538..295M}.  Alternatively, local thermal instability can develop if the $\alpha$-parameter is proportional to the magnetic Prandtl number \citep{2014MNRAS.441..681P, 2017arXiv170402485P}.  Work has gone into understanding these mechanisms analytically and with simple numerical models, but their viability in a global MHD accretion disk is untested.  Numerical models must rigorously test these mechanisms to determine which are the most feasible and to tie them to observational signatures that could originate at the transition.  The latter goal could provide a fruitful avenue for informing the large, time-domain surveys scheduled for the near future.

The second unique insight that comes from studying the MHD truncated disk is the interaction with the magnetic dynamo; specifically, that the dynamo was disrupted in the truncated disk region and that the outflow strengthened the field surrounding the cooler disk body.  Some form of low-frequency magnetic dynamo is universally observed in three-dimensional MHD accretion disk models.  Usually, the dynamo appears as quasi-periodic oscillations of the large-scale magnetic field in a standard thin accretion disk.  However, there is a developing narrative that certain conditions can hinder the organization of the large-scale field by the dynamo and impede the regular strengthening of the dynamo cycle.  A few instances from numerical models include quenching in a magnetically dominated disk \citep{2013ApJ...767...30B, 2016MNRAS.457..857S} and the mixing of field from hydrodynamic convection \citep{2017MNRAS.467.2625C}.  Global simulations of super-Eddington flows that include radiative effects find that the dynamo period remains regular, but is slower when than what it should normal be \citep{2014ApJ...796..106J}.  Nonphysical causes like resolution effects from sensitivities to the simulation domain aspect ratio have also been shown to quench the dynamo \citep{2017arXiv170408636W}.

The sporadic dynamo oscillations in the inner regions of the truncated disk provide another example of how the dynamo behavior can be modified.  Interior to the transition the magnetic field might have been expected to scale with the increase in gas pressure.  Instead, we find that $\beta$ increases and there is a steep drop in the effective $\alpha$-parameter to $\alpha=0.02$, or $\approx35\%$ of its pretransition value.  The inability of the magnetic field to adjust and increase to maintain the higher $\alpha$-parameter is most likely due to the outflow launched from the gas transitioning between the two cooling branches in the transition zone.  This modifies the flow dynamics and disrupts the organization of the large-scale field.

Oddly enough, the same outflow that suppresses the dynamo in the inner disk produces a thin layer of enhanced magnetic field that envelopes the transition zone.  This occurs as the outflowing gas stretches the magnetic field when it flows over the tapered disk body, which converts kinetic energy into magnetic energy.  In addition to demonstrating a secondary pathway that can amplify field, this feature is interesting because of its observational prospects.  In particular, it could appear as a quasiperiodic oscillation (QPO).  QPO phenomenology is rich and complex, but generally low-frequency QPOs appear in the PSDs of hard-state BHB light curves as strong and narrow peaks whose central frequency can evolve with the black hole outburst.  The magnetic dynamo has been suggested as a driver of this peculiar behavior \citep{2011ApJ...736..107O, 2015ApJ...809..118B} and our results offer an extension to this model that may bolster its theoretical underpinnings.  Specifically, the additional field strengthening occurs at the innermost part of the truncated disk, which is already the most luminous region of the disk.  Given that this emission is quite localized, it will preferentially weight a small range of frequencies and could explain the high quality factors that are typically observed.  Frequency evolution in the signal is naturally accounted for in this model because the QPO frequency is tied to the orbital and dynamo frequencies, and will therefore change as the secular changes in the mass accretion rate change the truncation radius of the disk. 

Headway can be made in more accurately modeling the dynamics of truncated accretion disks by including a more realistic radiation prescription, which we have purposefully treated with a simple optically thin approximation.  The accretion flow and transition zone behavior could be altered from the model analyzed here because radiation pressure could modify the flow dynamics and the thermal transition could evolve differently.  Our aim was to conduct a generalized investigation that is applicable to black holes across the mass scale.  Uniquely targeting truncated accretion disks in either outbursting stellar mass black holes in the hard-state or LLAGNs requires accounting for the detailed thermodynamics, which are quite different between the two regimes.  The ability of the radiation field to impart momentum to the gas and help launch an outflow \citep{2000ApJ...538..684P} is likely to be a significant contributor to the thermal acceleration we find here.  In AGNs, line driving could also lead to truncation \citep{2014MNRAS.438.3024L} and iron opacity may be important to the disk structure \citep{2016ApJ...827...10J}.

\section{Conclusions}
\label{sec-conclusion}

We investigate the dynamics of a truncated accretion disk around a black hole using a well-resolved, semi-global MHD model.  An \emph{ad hoc} bistable cooling function is implemented to mimic the radiative transitions thought to be responsible for the distinct spectral states in stellar mass and supermassive black hole systems.  This cooling function produces a truncated accretion disk where the outer disk cools efficiently and resembles the standard Shakura $\&$ Sunyaev thin disk.  Within the truncation the gas is hot and diffuse, and the disk thickens.  Between these two regions lies a ``transition zone" from $r=130-200\:r_g$ where the cool, Keplerian gas disk tapers into the hot flow.  A thermally driven outflow is launched from this region, slightly depleting the accretion flow.  As the outflow buoyantly rises along the disk body, we find it amplifies the magnetic field on the edge of the main disk body.  The turbulent intensity ($\overline{\delta}$) is boosted in the transition zone as a result of the transitioning gas and the Maxwell stress is increased from the outflow; in stark contrast to the inner disk where $\overline{\delta}$ is at a minimum.  In addition to the gas temperature and density, we find that the inner accretion flow of the truncated disk is differentiated from the outer cool disk in other ways as well.  In particular, the magnetic dynamo is suppressed and the effective $\alpha$-parameter is greatly diminished.  Despite these changes between the inner and outer disk, hot gas is able to efficiently accrete throughout the duration of the simulation.  We do not find that the accretion is inhibited by the presence of the truncation, with the exception of a slight siphoning of gas by the outflow.  The accretion rate through the inner simulation domain is essentially set by the feeding from the outer disk with the truncated region filling the role of an intermediary between the outer Shakura $\&$ Sunyaev-like thin disk and the black hole.

Two aspects of the simulation are worth highlighting because they are pertinent to the interpretation of spectra from systems with a truncated accretion disk: the effectiveness of the hot phase gas to fill the volume of the coronal regions above the cold, thin disk and the influence of turbulent fluctuations in introducing a radial taper.  In this simulation, the upper atmosphere of the disk is full of hot, radiatively inefficient gas and the density is greater because of the mass loaded outflow.  A large effort has gone into constraining the coronal geometry empirically and models typically assume the hot gas lies within the truncation.  Our results, in conjunction with prior studies, suggest it may not be the case that hot gas is only confined to a spherical region with a size set by the truncation radius.  A configuration where the hot gas freely fills the volume around the black hole above and below the cold, thin disk may be more realistic.  Additionally, consensus has been reached that MRI driven turbulence is the chief mediator of the angular momentum transport and disk heating.  Incorporating the modification of the truncation from a sharp interface to a more gradual radial dependence with a vertical temperature stratification because of the turbulent fluctuations may lead to more accurate measurements of truncation radii in these systems.  Since the truncation zone accounts for an appreciable portion of the disk, it is unclear where iron reflection measurements would place the truncation edge in such a system.  Adding sophistication to a well-resolved, long-duration, global MHD truncated accretion disk model by including more accurate radiative physics will help alleviate outstanding uncertainty and advance the understanding of the dynamics of truncated accretion disks.

\acknowledgements

The authors thank the anonymous referee for useful comments that helped strengthen and clarify the paper.  JDH thanks support by NASA under the NASA Earth and Space Science Fellowship program.  CSR thanks support from a Sackler Fellowship (hosted by the Institute of Astronomy, Cambridge) and NASA under grant NNX15AC40G.  The authors acknowledge the University of Maryland supercomputing resources (http://hpcc.umd.edu), including the Deepthought2, Deepthought, and MARCC/Bluecrab clusters, made available for conducting the research reported in this paper.

\end{document}